\documentclass[11pt]{article}

\usepackage{color}
\usepackage{graphicx}
\usepackage{graphics}
\usepackage{amscd}
\usepackage{amsmath}
\usepackage{amsfonts}
\usepackage{amssymb}
\usepackage{cite}

\DeclareMathOperator{\wt}{wt_H}

\newcommand{\bF}{ {\mathbb F}}
\newcommand{\PG}{ {\mathrm{PG} }}

\newcommand{\C}{ {\mathcal C}}

\newcommand{\bc}{ {\mathbf{c}}}
\newcommand{\bx}{ {\mathbf{x}}}

\newcommand{\tr}{ {\mathrm{Tr}}}

\newtheorem{theorem}{Theorem}[section]

\newtheorem{corollary}[theorem]{Corollary}

\newtheorem{example}[theorem]{Example}

\newtheorem{lemma}[theorem]{Lemma}

\newtheorem{remark}[theorem]{Remark}

\begin{document}

\title{ Some punctured codes of several families of binary linear codes }

\author{Xiaoqiang Wang,\,\,\, Dabin Zheng{\thanks{Corresponding author.
\newline \indent ~~Xiaoqiang Wang and Dabin Zheng are with the Hubei Key Laboratory of Applied Mathematics, Faculty of Mathematics and Statistics, Hubei University, Wuhan 430062, China (E-mail: waxiqq@163.com, dzheng@hubu.edu.cn).
\newline \indent ~~Cunsheng Ding is with the Department of Computer Science and Engineering, The Hong Kong University of Science and Technology, Clear Water Bay, Kowloon, Hong Kong, China (E-mail: cding@ust.hk).},\,\,\, Cunsheng Ding }}


\date{
}
\maketitle

\leftskip 0.8in
\rightskip 0.8in
\noindent {\bf Abstract.} Two general constructions of linear codes with functions over finite fields have been extensively studied in the literature. The first one is given by
$\mathcal{C}(f)=\left\{ {\rm Tr}(af(x)+bx)_{x \in \mathbb{F}_{q^m}^*}: a,b \in \mathbb{F}_{q^m} \right\}$, where $q$ is a prime power, $\bF_{q^m}^*=\bF_{q^m} \setminus \{0\}$, $\tr$
is the trace function from $\bF_{q^m}$ to $\bF_q$,  and $f(x)$ is a function from $\mathbb{F}_{q^m}$ to $\mathbb{F}_{q^m}$ with $f(0)=0$. Almost bent functions, quadratic functions
and some monomials on $\bF_{2^m}$ were used in the first construction, and many families of binary linear codes with few weights were obtained in the literature. This paper studies
some punctured codes of these binary codes. Several families of binary linear codes with few weights and new parameters are obtained in this paper. Several families of distance-optimal
binary linear codes with new parameters are also produced in this paper.



\vskip 35pt

\leftskip 0.0in
\rightskip 0.0in

\section{{\bf Introduction of motivations, objectives, and methodology}}

Let $q$ be a prime power and $n$ be a positive integer. An $[n, k, d]$ code $\mathcal{C}$ over the finite field $\mathbb{F}_q$ is a $k$-dimensional linear subspace of $\mathbb{F}_q^n$ with minimum Hamming distance $d$. The dual code, denoted by $\C^\perp$, of $\C$ is defined by
$$
\C^\perp=\left\{\bx=(x_0, \ldots, x_{n-1}) \in \bF_q^n: \sum_{i=0}^{n-1} x_ic_i=0 \  \forall \  \bc=(c_0, \ldots, c_{n-1}) \in \C \right\}.
$$
The minimum distance of $\C^\perp$, denoted by $d^\perp$, is called the dual distance of $\C$.
$\C$ is called a projective code if its dual distance is at least $3$.
An $[n, k, d]$
code over $\bF_q$ is said to be \textit{distance-optimal} (respectively,
\textit{dimension-optimal} and \textit{length-optimal}) if there is no $[n, k, d' \geq d+1]$ (respectively,
$[n, k' \geq k+1, d]$ and $[n' \leq n-1, k, d]$) linear code over $\bF_q$. An optimal code
is a code that is length-optimal, or dimension-optimal, or distance-optimal, or meets a bound for linear codes.  A binary linear code $\mathcal{C}$ is called {\it self-complementary} if it contains the all-one vector.  Let $A_i$ denote the number of  codewords with Hamming weight $i$ in $\mathcal{C}$. The {\it weight enumerator} of $\mathcal{C}$ is defined by $1+A_1x+A_2x^2+\cdots+A_nx^n$. The {\it weight distribution} of $\mathcal{C}$ is defined by the sequence $(1, A_1, \cdots, A_n)$. If the number of nonzero $A_i$ in the sequence $(A_1, \cdots, A_n)$ is $t$, then  the code $\mathcal{C}$ is said to be a $t$-weight code. By the parameters of a code, we mean its length, dimension and minimum distance.

Coding theory has important applications in communications systems, data storage systems, consumer electronics, and cryptography.  In addition, coding theory
is closely related to many areas of mathematics, such as algebra, algebraic geometry, algebraic function fields, algebraic
number theory, association schemes, combinatorics, finite fields, finite geometry, graph theory, and group theory. These
are the major motivations of studying coding theory. Constructing linear codes with desired parameters and weight distributions has been an important task in the history of coding theory. Linear codes may be constructed directly with algebraic approaches, combinatorial approaches and other approaches. Alternatively, almost all linear codes over finite fields can be constructed from some known codes by the puncturing or shortening techniques.

Let $\C$ be an $[n, k, d]$ code over $\bF_q$, and let $T$ be a set of $t$ coordinate positions in $\C$. We puncture $\C$ by deleting all the coordinates in $T$ in each codeword of $\C$. The resulting
code is still linear and has length $n-t$, where $t=|T|$. We denote the punctured code by $\C^T$. Let $\C(T)$ be the set of codewords which are $0$ on $T$. Then $\C(T)$ is a subcode of
$\C$. We now puncture $\C(T)$ on $T$, and obtain a linear code over $\bF_q$ with length $n-t$, which is called a \emph{shortened code}\index{shortened code} of $\C$, and is denoted by $\C_T$.
The puncturing and shortening techniques are two very important tools for constructing new codes from old ones. It was shown that every projective linear code over $\bF_q$ (i.e., the minimum distance of the dual code is at least 3) is a punctured code of a Simplex code over $\bF_q$ and a shortened code of a Hamming code over $\bF_q$ \cite{LDT20}. These facts justify the importance of the Simplex codes and the Hamming codes as well as the puncturing and shortening techniques. Note that the Simplex codes are optimal with respect to the Griesmer bound. Since every projective code is a punctured Simplex code, a punctured code of an optimal linear code may have good or bad parameters. To obtain a very good punctured code $\C^T$ from a good or optimal linear code $\C$, one has to choose a proper set $T$ of coordinate positions in $\C$. This is the difficulty of using the puncturing technique to construct new linear codes with good parameters from old ones \cite{LDT20,XTD20}. In this paper, we will use the puncturing technique to construct new codes with interesting and new parameters from some old linear codes.

Linear codes with few weights have applications in secret sharing~\cite{A1998}, strongly regular graphs~\cite{CA1986}, association schemes~\cite{CA1984} and authentication codes~\cite{Ding2005}. In finite geometry,
hyperovals in the projective geometry $\PG(2, 2^m)$ are the same as $[2^m+2, 3, 2^m]$ MDS codes with two weights \cite[Chapter 12]{DingBK18}, maximal arcs in $\PG(2, 2^m)$ are the same as a special type of two-weight codes
\cite[Chapter 12]{DingBK18}, and ovoids in $\PG(3, q)$ are the same as a special type of two-weight codes \cite[Chapter 13]{DingBK18}. Many families of linear codes have been used to construct combinatorial $t$-designs \cite[Chapters 5--13]{DingBK18}. These are some of the motivations of studying linear codes with few weights in the literature. In the past two decades, a lot of progress on the construction of linear codes with few weights has been made. The reader is referred to ~\cite{Dingetal2007, Ding2015, Ding2016,DDing2014,HengYue2015,LiYueFu2017, LuoCaoetal2018,Mesnagerbc,TangLietal2016,Wangetal2015,Wang2015,Wangetal2016, Tang2018,Xiaetal2017,Tan2018,ZhouLietal2015} and the references therein for information.  One of the objectives of this paper is to construct binary linear codes with few weights.

Functions and linear codes are closely connected. In the literature two general constructions of linear codes with functions over finite fields have been intensively investigated \cite{Ding2016}.
The first construction is given by
\begin{eqnarray}\label{eqn-codeffunction}
\mathcal{C}(f)=\left\{ {\rm Tr}(af(x)+bx)_{x \in \mathbb{F}_{q^m}^*}\,:\, a,b \in \mathbb{F}_{q^m} \right\},
\end{eqnarray}
where $q$ is a prime power, $\bF_{q^m}^*=\bF_{q^m} \setminus \{0\}$, $\tr$  is the trace function from $\bF_{q^m}$ to $\bF_q$, and $f(x)$ is a function from $\mathbb{F}_{q^m}$ to $\mathbb{F}_{q^m}$ with $f(0)=0$. It is clear that $\mathcal{C}(f)$  is a linear code with length $q^m-1$ and dimension at most~$2m$. If $f(x)$ is a monomial, then $\C(f)$ is permutation-equivalent to a cyclic code \cite{CDY05}.
This general construction has a long history and its importance is supported by Delsarte's Theorem \cite{Delsarte1975}.
The weight distribution of $\mathcal{C}(f)$ is closely related to the value distributions of certain exponential sums,
and is difficult to settle in general. In order to determine the weight distribution of $\C(f)$, people usually choose $f(x)$ to be a special function such as  a quadratic function, PN function, and APN function. Many good and optimal linear codes have been obtained with
this construction. This is also a main method for constructing linear codes with few weights.
The reader is referred to, for example,   \cite{CDY05,Hollmann,Fengluo2007,LuoFeng2008,Lietal2014,
Mesnagerbc,Wang2015, YCD06} for information.

The second general construction of linear codes is described as follows \cite{Dingetal2007,Wolfmann75}.   Let $D=\{d_1, d_2, \cdots, d_n\} \subset \bF_{q^m}^*$ be a multiset. Define a linear code
\begin{equation*}
\C_D = \left\{ \left( {\rm Tr}( xd_1), {\rm Tr}( xd_2), \cdots, {\rm Tr}(xd_n) \right) : x\in \bF_{q^m} \right\},
\end{equation*}
where $q$ is a prime power, $\tr$  is the trace function from $\bF_{q^m}$ to $\bF_q$.
The code $\C_D$ over $\bF_q$ has length $n$ and dimension at most $m$, where $D$ is called the defining set of $\C_D$. This construction is fundamental in the sense that every linear code over $\bF_q$
can be expressed as $\C_D$ for some positive integer $m$ and some subset $D$ of $\bF_{q^m}$  \cite{HWW20,Xiang16}.
It is known that this construction is equivalent to the generator matrix construction of linear codes.
The code $\mathcal{C}_D$ may have good parameters if the defining set is properly chosen.
With the second general construction, many good linear codes with few weights have been constructed~\cite{Ding2015,DingLietal2015,DDing2015,HengYue2015,Li2020,LiYueFu2017,HengYueLi2016,LuoCaoetal2018,
Mesnagerbc, Wangetal2016,Wangetal2015}.
With some variants of the second construction, interesting linear codes were obtained in \cite{Tang2017,LiYueFu2017,LIBaeFu2019}.

By the definition of the second construction above, $\C_{\bF_{q^m}^*}$ has parameters $[q^m-1, m, (q-1)q^{m-1}]$
and weight enumerator
$1+(q^m-1)z^{(q-1)q^{m-1}}$. If $D  \subset \bF_{q^m}^*$ does not contain repeated elements, let
$\bar{D}=\bF_{q^m}^* \setminus D$. In this case, we have $\C_D=(\C_{\bF_{q^m}^*})^{\bar{D}}$, where
the coordinate positions in $\C_{\bF_{q^m}^*}$ are indexed by the elements in $\bF_{q^m}^*$. This means
that $\C_D$ is in fact a punctured code of the one-weight code $\C_{\bF_{q^m}^*}$, which is a concatenation of
$(q-1)$ Simplex codes over $\bF_q$ with the same parameters. Hence, the second construction above is in fact
a puncture construction, and every projective linear code over $\bF_q$ is a punctured code of the
one-weight code $\C_{\bF_{q^m}^*}$.

Motivated by the power of the puncture technique and the first construction, in this paper we study some
punctured codes of several families of binary linear codes $\C(f)$ from special functions on $\bF_{2^m}$.
Specifically, we will study the following punctured codes.

Let $f$ be a function on $\bF_{2^m}$ with $f(0)=0$, and let $D=\{d_1, d_2, \cdots, d_n\}\subset \bF_{2^m}^*$ that
does not contain any repeated elements. Define $\bar{D}=\bF_{2^m}^* \setminus D$. In this paper, we will study the
punctured code
\begin{equation}\label{eq:cxxx}
\C(f)^{\bar{D}} = \left\{ \bc(a,b)= \left( {\rm Tr}( af(d_1)+b d_1), \cdots, {\rm Tr}( af(d_n)+b d_n) \right):  a,b\in \bF_{2^m} \right\},
\end{equation}
where $\tr$ is the trace function from $\bF_{2^m}$ to $\bF_2$ and the binary code $\C(f)$ was defined in (\ref{eqn-codeffunction}).  We call the set $D$ the \textit{position set} of the code $\C(f)^{\bar{D}}$, as we index the coordinate positions of the code $\C(f)$ with the elements in $\bF_{2^m}^*$. The dimension of $\C(f)^{\bar{D}}$ is at most $2m$. The two objectives of this paper are to obtain binary linear codes
$\C(f)^{\bar{D}}$ with new parameters and few weights and $(\C(f)^{\bar{D}})^\perp$ with new and good parameters. To this end, we have to select $f$ and the position set $D$ carefully.

Concretely, we first choose the position set to be
\begin{equation}\label{eq:cxxdx}
\begin{split}
D=\left\{ x\in \mathbb{F}_{2^m}^*\,: \, {\rm Tr}(\lambda f(x))=\nu \right\}
\end{split}
\end{equation}
and determine the weight distributions of $\C(f)^{\bar{D}}$, where $\nu \in \left\{ 0, 1\right\}$, $\lambda \in \mathbb{F}_{2^m}^*$ and $f(x)$ is an almost bent function from $\bF_{2^m}$ to itself.  We show that $\C(f)^{\bar{D}}$ is a five-weight code
if $\nu=0$ and a self-complementary six-weight code if $\nu=1$. Some of the codes $\C(f)^{\bar{D}}$  are optimal according to the tables of best codes known in \cite{Grassl2006}. The dual of $\C(f)^{\bar{D}}$
 is distance-optimal with respect to the sphere packing bound if  $\nu=1$.
We then present several classes of four-weight or six-weight linear codes by choosing $f(x)$ to be some special quadratic functions, and the position set to be the {\it support} of ${\rm Tr}(x)$, i.e.,
\begin{equation}\label{eqdd}
D=\left\{ x\in \mathbb{F}_{2^m}^*\,: \, {\rm Tr}(x)=1 \right\}.
 \end{equation}
Several families of complementary binary linear codes are obtained. The parameters of the duals of $\C(f)^{\bar{D}}$ are also determined and almost all of them are distance-optimal with respect to the sphere packing bound. Finally, we present several classes of binary linear codes with three weights, or five weights or six weights by selecting the position sets to be some cyclotomic classes. Some of the codes and their duals are distance-optimal. The parameters of most of the codes presented in this paper are new.

The rest of this paper is organized as follows. Section~\ref{sec-hubei2} introduces some preliminaries.
Section~\ref{sec-hubei3} investigates the weight distribution of the linear code $\C(f)^{\bar{D}}$ and the parameters of its dual,
where $f(x)$ is an almost bent function, $D=\left\{ x\in \mathbb{F}_{2^m}^* : {\rm Tr}(\lambda f(x))=\nu \right\}$, $\nu \in \left\{ 0, 1\right\}$ and $\lambda \in \mathbb{F}_{2^m}^*$. Section~\ref{sec-hubei4} determines the weight distribution of the linear code $\C(f)^{\bar{D}}$ and the parameters of its dual, where $f(x)$ is some special quadratic function  and $D=\left\{ x\in \mathbb{F}_{2^m}^* : {\rm Tr}(x)=1 \right\}$. Section~\ref{sec-hubei5} settles the weight distribution of the linear code $\C(f)^{\bar{D}}$ and the parameters of its dual, where $D$ is a cyclotomic class and $f$ is a monomial. Section \ref{sec:concluding} concludes this paper.

\section{{\bf Preliminaries}}\label{sec-hubei2}

In this section, we introduce some special functions on $\bF_{2^m}$, some exponential sums and some
basic results in coding theory, which will be used later in this paper.

\subsection{Notation used starting from now on}
Starting from now on, we assume $m\geq 4$ and adopt the following notation unless otherwise stated:
\begin{description}
\item{$\bullet$} $\bF_{2^m}$ is the finite field with $2^m$ elements and $\gamma$ is a primitive element of $\mathbb{F}_{2^m}$.
\item{$\bullet$} $\bF_{2^m}^*=\bF_{2^m} \setminus \{0\}$.
\item{$\bullet$} ${\rm Tr}(\cdot)$ is the absolute trace function from $\bF_{2^m}$ to $\bF_{2}$.
\item{$\bullet$} ${\rm Tr}_u^v(\cdot)$ is the trace function from $\bF_{2^v}$ to $\bF_{2^u}$, where $u,v$ are positive integers such that $u\,|\,v$.
\item{$\bullet$} $v_2(\cdot)$ is the 2-adic order function with $v_2(0)=\infty$.
\item{$\bullet$} $\wt(\bf c)$ denotes the Hamming weight of a vector $\bc$.
\item{$\bullet$} $d_{H}(\C)$ denotes the minimum distance of a linear code $\C$.
\end{description}

\subsection{AB and APN functions}

Let $f(x)$ be a function from $\mathbb{F}_{2^m}$ to $\mathbb{F}_{2^m}$. The {\it Walsh transform} of $f(x)$ at $(a,b) \in \mathbb{F}_{2^m}^2$ is defined as
\begin{equation}\label{Wfab}
W_f(a,b)=\sum_{x \in \mathbb{F}_{2^m}}(-1)^{{\rm Tr}(af(x)+b x)}.
\end{equation}
If $W_f(a,b)=0$ or $\pm 2^{\frac{m+1}{2}}$ for any pair $(a,b)\in \bF_{2^m}^2$ with $a\neq 0$, then $f(x)$ is called an {\it almost bent (AB) function}. Almost bent functions exist only for odd $m$.
Define
$$\delta_{f}(a,b)={\rm max}_{a \in \mathbb{F}_{2^m}^*, b \in \mathbb{F}_{2^m}}|\{ x \in \mathbb{F}_{2^m}\, : \, f(x+a)+f(x)=b\}|,$$
then $f(x)$ is called an {\it almost perfect nonlinear (APN) function} if $\delta_{f}(a,b)= 2$.

APN and AB functions have applications in coding theory, combinatorics, cryptography, finite geometry and sequence design.
Many good linear codes over finite fields have been constructed with APN and AB functions~\cite{CCZ98,Ding2015,Ding2016,Lietal2014,Mesnagerbc}. AB functions and APN functions have the following relationship.

\begin{lemma}\cite{Budaghyan2006}\label{lem:almost}
Let $\mathbb{F}_{2^m}$ be a finite field with $2^m$ elements. If $f(x)$ is an almost bent function over $\mathbb{F}_{2^m}$, then $f(x)$ is an almost perfect nonlinear function over $\mathbb{F}_{2^m}$.
\end{lemma}
The converse is not true for Lemma \ref{lem:almost}, as almost bent functions exist only for $m$ being odd while almost perfect nonlinear functions exist for $m$ being even too.

\subsection{Quadratic functions}

By identifying the finite field $\bF_{2^m}$ with the $m$-dimensional vector space $\bF_{2}^m$ over $\bF_2$,
a function~$f$ from $\bF_{2^m}$ to $\bF_2$ can be viewed as an $m$-variable polynomial over $\bF_2$.
In the sequel, we fix a basis of $\bF_{2^m}$ over $\bF_{2}$ and identify $x\in \bF_{2^m}$ with a vector
$(x_1, x_2, \cdots, x_m)\in \bF_{2}^m$, a quadratic function over $\bF_{2}$ is of the form:
\begin{equation*}
Q(x_1, x_2, \cdots, x_m) = (x_1, x_2, \cdots, x_m) A (x_1, x_2, \cdots, x_m)^{T} ,
\end{equation*}
where $A=(a_{ij})_{m\times m}, \, a_{ij}\in \bF_{2}$, is an upper triangular matrix. The matrix $A+A^T$ is called an alternate matrix and its rank must be even~\cite{Wan03}.
By the theory of linear equations, the rank $r$ of the matrix $A+A^T$ is equal to the codimension of the $\bF_{2}$-linear subspace
\begin{equation}\label{eq:quadraticform1}
 V = \{ x\in \bF_{2^m}:  Q(x+z)+Q(x)+Q(z) = 0 \mbox{ for all } z\in \bF_{2^m} \},
\end{equation}
i.e. $r = m-\dim_{\bF_{2}}V$. Let $G(x)$ be a linear polynomial over $\mathbb{F}_{2^m}$, then
\begin{equation*}
\begin{split}
\left(\sum_{x \in \mathbb{F}_{2^m}}(-1)^{{\rm Tr}(Q(x)+G(x))}\right)^2&=\sum_{x \in \mathbb{F}_{2^m}}(-1)^{{\rm Tr}(Q(x)+G(x))}\sum_{y \in \mathbb{F}_{2^m}}(-1)^{{\rm Tr}(Q(y)+G(y))}\\
&=\sum_{x,y \in \mathbb{F}_{2^m}}(-1)^{{\rm Tr}(Q(x+y)+G(x+y)+Q(x)+G(x))}\\
&=\sum_{y \in \mathbb{F}_{2^m}}(-1)^{{\rm Tr}(Q(y)+G(y))}\sum_{x \in \mathbb{F}_{2^m}}(-1)^{{\rm Tr}(Q(x+y)+Q(x)+Q(y))}\\
&=2^m \cdot \sum_{y \in V}(-1)^{{\rm Tr}(Q(y)+G(y))},
\end{split}
\end{equation*}
where $V$ was defined in $(\ref{eq:quadraticform1})$. It is easy to check that
$${{\rm Tr}\left(Q(x+y) +G(x+y)\right)}={{\rm Tr}\left(Q(x) +G(x)\right)}+{{\rm Tr}\left(Q(y) +G(y)\right)}$$
for any $x,y \in V$. Then
\begin{equation}\label{eqWf2}
\begin{split}
\left(\sum_{x \in \mathbb{F}_{2^m}}(-1)^{{\rm Tr}(Q(x)+G(x))}\right)^2=\begin{cases}
2^{m+r}, & \text{if ${{\rm Tr}\left(Q(y) +G(y)\right)}=0$ for all $y \in V$,} \\
0, & \text{otherwise},
\end{cases}
\end{split}
\end{equation}
where $r$ is the rank of $Q(x)$ and $r = m-\dim_{\bF_{2}}V$. The following are some well known results about quadratic forms, which will be needed in this paper.

\begin{lemma}\cite{Coulter1998, Coulter2002}\label{lemeven1}
 Let $m$ and $ k$ be non-negative integers with $v_2(m) \leq v_2(k)$ and $a,b \in \mathbb{F}_{2^m}$ with $a\neq 0$. Let
 \begin{equation}\label{eqdfgg}
\begin{split}
S(a,b)=\sum_{x \in \mathbb{F}_{2^m}}(-1)^{{\rm Tr}\left(ax^{2^k+1}+bx\right)},
\end{split}
\end{equation}
 then
 the possible values of $S(a,b)$ are in the set $\{0, \pm 2^{\frac{m+\ell}{2}}\}$, where $\ell=\gcd(m,k)$.
\end{lemma}

\begin{lemma}\cite{Coulter1998, Coulter2002}\label{lemeven2}
 Let $m$ and $k$ be non-negative integers with $v_2(m) > v_2(k)$ and $a,b \in \mathbb{F}_{2^m}$ with $a\neq 0$. Let $S(a,b)$ be defined in (\ref{eqdfgg}). Then $S(a,b)=0$ unless the equation $a^{2^k}x^{2^{2k}}+ax+b^{2^k}=0$ is solvable. Let $\gamma$ be a primitive element of $\mathbb{F}_{2^m}$.  Let  $\ell=\gcd(m,k)$. Assume $a^{2^k}x^{2^{2k}}+ax+b^{2^k}=0$ is solvable. Then there are two possibilities as follows.
\begin{description}
\item{(i)} If $a\neq \gamma^{s\left(2^\ell+1\right)}$ for any integer $s$, then the equation has a unique solution $x_b$ for any $b \in \mathbb{F}_{2^m}$, and
    $$S(a,b)=(-1)^{\frac{m}{2\ell}-{\rm Tr}\left(ax_b^{2^k+1}\right)}2^{\frac{m}{2}}.$$

\item{(ii)} If $a= \gamma^{s\left(2^\ell+1\right)}$ for some integer $s$, then the equation is solvable if and only if ${\rm Tr}_{2\ell}^m\left(b\beta^{-s}\right)=0$, where $\beta \in \mathbb{F}_{2^m}^*$ is the unique element satisfying $\beta^{\frac{2^k+1}{2^\ell+1}}=\gamma$. In such case,
    $$ S(a,b)=-(-1)^{\frac{m}{2\ell}-{\rm Tr}\left(ax_b^{2^k+1}\right)}2^{\frac{m}{2}+\ell},$$
    where $x_b$ is a solution to $a^{2^k}x^{2^{2k}}+ax+b^{2^k}=0$.
\end{description}
\end{lemma}
\begin{lemma}\cite{Moisio2016} \label{lemeven3}
Let $\gamma$ be a primitive element of $\mathbb{F}_{2^m}$. Assume that $m=2sh$ and $\ell \, | \,(2^h+1)$. Then
\[\sum_{x \in \mathbb{F}_{2^m}}(-1)^{{\rm Tr}(\gamma^i x^\ell)}=\left\{ \begin{array}{lcl}
           (-1)^s2^{\frac{m}{2}}, & {\rm if} \,\,\,  i \not\equiv 0 \pmod \ell,\\
           (-1)^{s-1}(\ell-1)2^{\frac{m}{2}},
             & {\rm if} \,\,\,  i \equiv 0 \pmod \ell. \end{array}  \right.\]
\end{lemma}

\begin{lemma}\cite{LuoFeng2010}\label{lemeveeen3}
Let $\ell=\gcd(\frac{m}{2},k)$ and $\ell'=\gcd(\frac{m}{2}+k,2k)$. Let
$$S_1(a,b)=(-1)^{{\rm Tr}\left(ax^{2^k+1}+bx^{2^\frac{m}{2}+1}\right)}.$$ If $\ell'=2\ell$ and $(a,b)$ runs over $\mathbb{F}_{2^m} \times  \mathbb{F}_{2^{\frac{m}{2}}}$,  then
\begin{equation*}
\begin{split}
S_1(a,b)=\begin{cases}
2^{m},& {\rm occuring}\,\,\,  1\,\, {\rm time},  \\
-2^{\frac{m}{2}},& {\rm occuring}\,\,\,  \frac{2^{3k}(2^{\frac{m}{2}}-1)(2^m-2^{m-2k}-2^{m-3k}+2^{\frac{m}{2}}-2^{\frac{m}{2}-k}+1}{(2^k+1)(2^{2k}-1)}\,\, {\rm times},  \\
2^{\frac{m}{2}+k},& {\rm occuring}\,\,\, \frac{2^k(2^m-1)(2^m-2^{m-\ell}+2^{m-2\ell}+1)}{(2^k+1)^2} \,\, {\rm times},  \\
-2^{\frac{m}{2}+2k},& {\rm occuring}\,\,\, \frac{(2^{\frac{m}{2}-\ell}-1)(2^m-1)}{(2^k+1)(2^{2k}-1)}\,\,{\rm times}.
\end{cases}
\end{split}
\end{equation*}
\end{lemma}

\subsection{Pless power moments and the sphere packing bound}

To study the parameters of the duals of the punctured binary codes $\C(f)^{\bar{D}}$, we need the Pless power moments
of linear codes. Let $\mathcal{C}$ be a binary $[n, k]$ code, and denote its dual by $\mathcal{C}^{\perp}$. Let
 $A_i$ and $A^{\perp}_i$ be the number of codewords of weight $i$ in $\mathcal{C}$ and $\mathcal{C}^{\perp}$, respectively.
 The first five Pless power moments are the following~\cite[p. 131]{MacWilliam1997}:
\begin{equation*}
\begin{split}
&\sum_{i=0}^nA_i=2^k;\\
&\sum_{i=0}^niA_i=2^{k-1}(n-A_1^{\perp});\\
&\sum_{i=0}^ni^2A_i=2^{k-2}[n(n+1)-2nA_1^{\perp}+2A_2^{\perp}];\\
&\sum_{i=0}^ni^3A_i=2^{k-3}[n^2(n+3)-(3n^2+3n-2)A_1^{\perp}+6nA_2^{\perp}-6A_3^{\perp}];\\
&\sum_{i=0}^ni^4A_i=2^{k-4}[n(n+1)(n^2+5n-2)-4n(n^2+3n-2)A_1^{\perp}+4(3n^2+3n-4)A_2^{\perp}-24nA_3^{\perp}+24A_4^{\perp}].\\
 \end{split}
\end{equation*}
If $A_1^{\perp}=A_2^{\perp}=A_3^{\perp}=A_4^{\perp}=0$, then the sixth Pless power moment becomes the following:
\begin{equation*}
 \sum_{i=0}^ni^5A_i=2^{k-5}\cdot n^5+5\cdot 2^{k-4} \cdot n^4+15\cdot 2^{k-5} \cdot n^3-5 \cdot 2^{k-4} \cdot n^2-A_5^{\perp} \cdot 2^{k-5} \cdot 120.
\end{equation*}

We will need the following bound for binary linear codes later.

\begin{lemma}[The sphere packing bound]
Let $\mathcal{C}$ be an $[n, k,d]$ binary code. Then
$$2^n\geq 2^k\sum_{i=0}^{\lfloor \frac{d-1}{2} \rfloor}\left(
\begin{array}{cccc}
   n  \\
     i  \\
\end{array}
\right).$$
\end{lemma}

\section{Some punctured codes of the binary codes from almost bent functions}\label{sec-hubei3}

Recall the code $\C(f)$ defined in (\ref{eqn-codeffunction}). When $q=2$ and $f(x)=x^{2^h+1}$ with $\gcd(h, m)=1$ and $m$
being odd, the parameters and weight distribution of the binary code $\C(f)$ were settled in \cite{Kasami661,Kasami662}.
 When $q=2$, $m$ is odd and $f(x)$ is an almost bent function on $\bF_{2^m}$, the parameters and weight distribution of the binary code $\C(f)$ were settled in \cite{CCZ98}.  The binary code $\C(f)$  has parameters $[2^m-1, 2m, 2^{m-1}-2^{(m-1)/2}]$
and three nonzero weights  \cite{CCZ98}.
Let $\C(f)^{\bar{D}}$ be the binary punctured code defined in (\ref{eq:cxxx}) with position set $D$ in (\ref{eq:cxxdx}), where $f(x)$ is an almost bent function from $\bF_{2^m}$ to itself.
In this section, we investigate the weight distribution of the punctured code $\C(f)^{\bar{D}}$ and the parameters of its dual. We first give the length of the linear code $\C(f)^{\bar{D}}$ in the following
lemma.

\begin{lemma}\label{eqadsd}
Let $\C(f)^{\bar{D}}$ be the linear code defined in (\ref{eq:cxxx}) with the position set $D$ in (\ref{eq:cxxdx}), where $f(x)$ is an almost bent function from $\mathbb{F}_{2^m}$ to itself. Then the length $n$ of $\C(f)^{\bar{D}}$ is
\begin{equation*}
\begin{split}
n=|D|=\begin{cases}
2^{m-1}-(-1)^\nu2^{\frac{m-1}{2}}-1+\nu, & {\rm if} \,\, \,W_f(\lambda,0)=-2^{\frac{m+1}{2}},  \\
2^{m-1}+(-1)^\nu2^{\frac{m-1}{2}}-1+\nu, & {\rm if} \,\, \,W_f(\lambda,0)=2^{\frac{m+1}{2}},  \\
2^{m-1}-1+\nu, & {\rm if} \,\, \, W_f(\lambda,0)=0,
\end{cases}
\end{split}
\end{equation*}
where $W_f(\lambda,0)$ was defined in (\ref{Wfab}) and $\nu \in \{0,1\}$.
\end{lemma}

In order to apply the Pless power moments to determine the multiplicity of each Hamming weight of $\C(f)^{\bar{D}}$,
we need to investigate the minimum Hamming distance of its dual.

\begin{lemma}\label{lemdis}
Let $\C(f)^{\bar{D}}$ be the linear code defined in (\ref{eq:cxxx}) with the position set $D$ in (\ref{eq:cxxdx}),
where $f(x)$ is an almost bent function from $\mathbb{F}_{2^m}$ to itself. Then the dual distance is lower bounded by
\begin{equation*}
\begin{split}
d_H \left( \left(\C(f)^{\bar{D}} \right)^\perp \right)\geq
\begin{cases}
5,& {\rm if} \,\,\,  \nu=0,  \\
6, & {\rm if} \,\,\, \nu=1. \\
\end{cases}
\end{split}
\end{equation*}
\end{lemma}
{\it Proof.} It is easy to see $ d_H\left( (\C(f)^{\bar{D}})^\perp \right)  \geq 3$ from the definition of $\C(f)^{\bar{D}}$. Next, we show that
$d_H\left( (\C(f)^{\bar{D}})^\perp \right)   \neq 4$. The case of
$d_H\left( (\C(f)^{\bar{D}})^\perp \right)  \neq 3$ can be shown similarly, and we omit the details of the proof.

If $d_H\left( (\C(f)^{\bar{D}})^\perp \right)  = 4$, then there are four pairwise-distinct elements $x_1$, $x_2$, $x_3$ and $x_4$ in $\bF_{2^m}^*$ such that
\begin{equation*}
\begin{split}
\begin{cases}
{\rm Tr}(\lambda f(x_1))={\rm Tr}(\lambda f(x_2))={\rm Tr}(\lambda f(x_3))={\rm Tr}(\lambda f(x_4))=\nu,\\
a(x_1+x_2+x_3+x_4)+b(f(x_1)+f(x_2)+f(x_3)+f(x_4))=0
\end{cases}
\end{split}
\end{equation*}
for any $a,b \in \mathbb{F}_{2^m}$. Then,
\begin{equation}\label{eq:xxyy}
\begin{split}
\begin{cases}
{\rm Tr}(\lambda f(x_1))={\rm Tr}(\lambda f(x_2))={\rm Tr}(\lambda f(x_3))={\rm Tr}(\lambda f(x_4))=\nu,\\
x_1+x_2+x_3+x_4=0,\\
f(x_1)+f(x_2)+f(x_3)+f(x_4)=0.
\end{cases}
\end{split}
\end{equation}
The second and third equations in (\ref{eq:xxyy}) can be rewritten as
\begin{equation*}
\begin{split}
\begin{cases}
x_1+x_2=\alpha\,\,\text{and}\,\, x_3+x_4=\alpha,\\
f(x_1)+f(x_2)=\beta\,\,\text{and}\,\,f(x_3)+f(x_4)=\beta,
\end{cases}
\end{split}
\end{equation*}
where $\alpha, \beta\in \mathbb{F}_{2^m}$ with $\alpha\neq 0$. Hence, there are four different elements $x_1$, $x_1+\alpha$, $x_3$ and $x_3+\alpha$ satisfying the equation
$f(x)+f(x+\alpha)=\beta$. This contradicts Lemma \ref{lem:almost}, as $f(x)$ is an almost perfect nonlinear function. Therefore,
$d_H\left( (\C(f)^{\bar{D}})^\perp \right)  \geq 5$.

If $\nu=1$ and  $ d_H\left( (\C(f)^{\bar{D}})^\perp \right)  = 5$, there are five pairwise-distinct elements $x_1,x_2,x_3,x_4,x_5$
in $\bF_{2^m}^*$ such that  $f(x_1)+f(x_2)+f(x_3)+f(x_4)+f(x_5)=0$ by the definition of $\C(f)^{\bar{D}}$, then ${\rm Tr}(\lambda(f(x_1)+f(x_2)+f(x_3)+f(x_4)+f(x_5)))=0$, which is contradictory  to
${\rm Tr}(\lambda f(x_1))={\rm Tr}(\lambda f(x_2))={\rm Tr}(\lambda f(x_3))={\rm Tr}(\lambda f(x_4))={\rm Tr}(\lambda f(x_5))=1.$
Hence,
\begin{equation*}
\begin{split}
d_H\left( (\C(f)^{\bar{D}})^\perp \right)
\geq\begin{cases}
5,& \text{if  $\nu=0$,}  \\
6, & \text{if $\nu=1$.}
\end{cases}
\end{split}
\end{equation*}
This completes the proof of this lemma. $\square$

We now give the weight distribution of the binary code $\C(f)^{\bar{D}}$ and the parameters of its dual as follows.

\begin{theorem}\label{Theorem1}
Let $\C(f)^{\bar{D}}$ be the linear code defined in (\ref{eq:cxxx}) with the position set $D$ in (\ref{eq:cxxdx}), where $f(x)$ is an almost bent function from $\mathbb{F}_{2^m}$ to itself. Then the following statements hold.
\begin{description}
\item{(1)} If $\nu=0$, then $\C(f)^{\bar{D}}$ is an $[n,2m-1,\frac{n+1}{2}-2^{\frac{m-3}{2}}]$ code with the weight distribution in Table~\ref{Table1}, where $n$ was given in Lemma \ref{eqadsd}. Its dual has parameters $[n,n-2m+1,5]$.
  \begin{table}[h]
\caption{\rm Weight distribution of the code $\C(f)^{\bar{D}}$ for $\nu=0$ in Theorem \ref{Theorem1} }\label{Table1}
\begin{center}
\begin{tabular}{cccc}\hline
Weight      & Multiplicity \\ \hline
$0$ &  $1$  \\ \hline
  $\frac{n+1}{2}$ & $2^{2m-1}-(n+1)^42^{-2m}+5(n+1)^22^{-m-1}-5(n+1)2^{m-2}+\frac{3}{2}n^2+2n-\frac{1}{2}$  \\   \hline
 $\frac{n+1}{2}\pm2^{\frac{m-1}{2}}$ & $\begin{array}{c}\pm \frac{1}{6}\big((n+1)^32^{\frac{1-3m}{2}}-(3n+1)2^{\frac{m-1}{2}}-(n+1)2^{-\frac{m+1}{2}}+2^{\frac{3m-3}{2}}\big)-\\
 \frac{1}{6}(n+1)^42^{-2m}+\frac{1}{6}(n+1)^22^{-m-1}-\frac{1}{6}(n+1)2^{m-2}+\frac{1}{4}n^2+\frac{1}{3}n+\frac{1}{12}\end{array}$  \\   \hline
$\frac{n+1}{2}\pm2^{\frac{m-3}{2}}$ & $\begin{array}{c}\pm\frac{1}{6}\big(-(n+1)^32^{\frac{3-3m}{2}}+(n+1)2^{\frac{5-m}{2}}+2^{\frac{m+1}{2}}-2^{\frac{3+3m}{2}}+6n\cdot 2^{\frac{m-1}{2}}\big)+2^{2-2m}\cdot n^2+\\
\frac{1}{3}(n^4+4n^3+4n+1)2^{1-2m}-\frac{1}{3}(n+1)^22^{2-m}+\frac{1}{3}(n+1)2^{1+m}-n^2-\frac{4}{3}n-\frac{1}{3}\end{array}$  \\   \hline
\end{tabular}
\end{center}
\end{table}
\item{(2)} If $\nu=1$, then $\C(f)^{\bar{D}}$ is an $[n,2m,\frac{n}{2}-2^{\frac{m-3}{2}}]$ code with the weight distribution  in Table~\ref{Table2}, where $n$ was given in Lemma \ref{eqadsd}. Its dual has parameters $[n,n-2m,6]$, and is distance-optimal with respect to the sphere packing bound.
\begin{table}[h]
\caption{\rm Weight distribution of the code $\C(f)^{\bar{D}}$ for $\nu=1$ in Theorem \ref{Theorem1} }\label{Table2}
\begin{center}
\begin{tabular}{cccc}\hline
Weight      & Multiplicity \\\hline
$0$ &  $1$ \\   \hline
  $\frac{n}{2}$ & $2^{2m}-5 n\cdot 2^{m-1}+5n^2\cdot 2^{-m}-2^{1-2m}n^4+3n^2-2n-2$  \\   \hline
 $\frac{n}{2}\pm2^{\frac{m-1}{2}}$ & $-\frac{1}{3}n^42^{-2m}+\frac{1}{6}(2^{-m}n^2+3n^2-2^{m-1}n-2n)$  \\   \hline
$\frac{n}{2}\pm2^{\frac{m-3}{2}}$ & $\frac{4n}{3}(2^{-2m}n^3-2^{1-m}n-\frac{3n}{2}+2^m+1)$  \\   \hline
$n$ &  $1$ \\   \hline
\end{tabular}
\end{center}
\end{table}
\end{description}
\end{theorem}

{\it Proof.}  It follows from (\ref{eq:cxxx}) that the Hamming weight of the codeword $\bc(a,b)$ in $\C(f)^{\bar{D}}$ is given~by
\begin{equation}\label{Nab01400}
\begin{split}
{\rm wt_H}(\mathbf{c}(a,b))&=|D|-\left|\left\{x\in D:\,\,  {\rm Tr}\left(af(x)+bx\right)=0 \right\}\right|\\
&=\frac{|D|}{2}-\frac{1}{2}\sum_{x \in D}(-1)^{ {\rm Tr}(af(x)+bx)}\\
&=\frac{|D|}{2}-\frac{1}{2}\sum_{x \in \mathbb{F}_{2^m}\setminus\{0\}}\left(\frac{1}{2}\sum_{y \in \mathbb{F}_{2}}(-1)^{y{({\rm Tr}(\lambda f(x))-\nu)}}\right)(-1)^{{\rm Tr}\left(af(x)+b x\right)}\\
&=\frac{|D|}{2}-\frac{1}{4}\sum_{x \in \mathbb{F}_{2^m}}\left(\sum_{y \in \mathbb{F}_{2}}(-1)^{y{({\rm Tr}(\lambda f(x))-\nu)}}\right)(-1)^{{\rm Tr}(af(x)+b x)}+\frac{1}{4}\sum_{y \in \mathbb{F}_{2}}(-1)^{y\nu}\\
&=\frac{|D|}{2}-\frac{1}{4}\sum_{x \in \mathbb{F}_{2^m}}\left(1+(-1)^{({\rm Tr}(\lambda f(x))-\nu)}\right)(-1)^{{\rm Tr}(af(x)+b x)}+\frac{1}{4}\sum_{y \in \mathbb{F}_{2}}(-1)^{y\nu}\\
&=\frac{|D|}{2}-\frac{1}{4}\sum_{x \in \mathbb{F}_{2^m}}(-1)^{{\rm Tr}(af(x)+b x)}-(-1)^{\nu}\sum_{x \in \mathbb{F}_{2^m}}(-1)^{{\rm Tr}((\lambda+a)f(x)+b x)}+\frac{1}{4}\sum_{y \in \mathbb{F}_{2}}(-1)^{zy\nu}\\
&=\frac{|D|}{2}-\frac{1}{4}W_f(a,b)-\frac{(-1)^{\nu}}{4}W_f(a+\lambda,b)+\frac{1}{4}\sum_{y \in \mathbb{F}_{2}}(-1)^{z_0\nu},
\end{split}
\end{equation}
where $W_f(a,b)$ was defined in (\ref{Wfab}). By the definition of almost bent functions, for any $(a,b) \in \mathbb{F}_{2^m}^2\setminus \{(0,0)\}$, we know that $W_f(a,b) \in \{0, \pm 2^{\frac{m+1}{2}}\}$. So,
\begin{equation}\label{eq:pm}
\frac{1}{4}\left(W_f(a,b) \pm W_f(a+\lambda,b)\right) \in \left\{0, \pm 2^{\frac{m-1}{2}}, \pm 2^{\frac{m-3}{2}}\right\}
\end{equation}
for any $(a,b)\in \mathbb{F}_{2^m}^2\setminus\{(0,0),(\lambda,0)\}$. In the following, we prove this theorem case by case.

\noindent {\bf Case 1:} $\nu=0$, i.e., $D=\{x \in \mathbb{F}_{2^m}^{*}: {\rm Tr}(\lambda f(x))=0\}$. By (\ref{Nab01400}) and (\ref{eq:pm}), when $(a,b)$ runs over $\mathbb{F}_{2^m}^2\setminus\{(0,0),(\lambda,0)\}$, the possible values of $\wt(\mathbf{c}(a,b))$ are
$$\frac{n+1}{2},\,\,\frac{n+1}{2} \pm 2^{\frac{m-1}{2}},\,\, \text{and}\,\, \frac{n+1}{2} \pm 2^{\frac{m-3}{2}},$$
where $n$ was given in Lemma \ref{eqadsd}. It is easy to see that $\wt(\mathbf{c}(a,b))=0$ if and only if $(a,b)=(0,0)$ or
$(a,b)=(\lambda,0)$. So, the dimension of $\C(f)^{\bar{D}}$ is $2m-1$.

Denote $w_1=\frac{n+1}{2}$, $w_2=\frac{n+1}{2}+2^{\frac{m-1}{2}}$, $w_3=\frac{n+1}{2}-2^{\frac{m-1}{2}}$, $w_4=\frac{n+1}{2}+2^{\frac{m-3}{2}}$ and $w_5=\frac{n+1}{2}-2^{\frac{m-3}{2}}$. Let $A_{w_i}$ be the number of the codewords with weight $w_i$ in $\C(f)^{\bar{D}}$.
By Lemma \ref{lemdis}, we know that $A_1^{\perp}=A_2^{\perp}=A_3^{\perp}=A_4^{\perp}=0$. From the first five Pless power
moments, we have the following system of equations:
\begin{equation*}
\begin{split}
\begin{cases}
\sum_{i=1}^5A_{w_i}=2^{2m-1}-1;\\
\sum_{i=1}^5w_iA_{w_i}=2^{2m-2}n;\\
\sum_{i=1}^5w_i^2A_{w_i}=2^{2m-3}n(n+1);\\
\sum_{i=1}^5w_i^3A_{w_i}=2^{2m-4}n^2(n+3);\\
\sum_{i=1}^5w_i^4A_{w_i}=2^{2m-5}n(n+1)(n^2+5n-2).\\
\end{cases}
\end{split}
\end{equation*}
Solving this system of equations, we obtain the desired values of $A_{w_1}$, $A_{w_2}$, $A_{w_3}$, $A_{w_4}$ and $A_{w_5}$ in Table \ref{Table1}.

We now determine the parameters of the dual of $\C(f)^{\bar{D}}$. We consider only the case $n=2^{m-1}-1$, i.e., the value of $W_f(\lambda,0)$ is zero. The other two cases can be shown similarly. Substituting the value of $n=2^{m-1}-1$ in Table \ref{Table1}, we obtain that $A_{w_1}=3\cdot2^{2m-4}+2^{m-3}-1$, $A_{w_2}=2^{2m-5}-2^{\frac{3m-7}{2}}+2^{\frac{m-5}{2}}-2^{m-4}$,
$A_{w_3}=2^{2m-5}+2^{\frac{3m-7}{2}}-2^{\frac{m-5}{2}}-2^{m-4}$, $A_{w_4}=2^{2m-3}-2^{\frac{3m-5}{2}}$ and $A_{w_5}=2^{2m-3}+2^{\frac{3m-5}{2}}$.
By Lemma \ref{lemdis}, $A_1^{\perp}=A_2^{\perp}=A_3^{\perp}=A_4^{\perp}=0$. Then from the sixth Pless power moment, we have
\begin{equation*}
\begin{split}
\sum_{i=1}^5w_i^5A_{w_i}&=2^{2m-6}\cdot (2^{m-1}-1)^5+5\cdot 2^{2m-5} \cdot (2^{m-1}-1)^4\\
&+15\cdot 2^{2m-6} \cdot (2^{m-1}-1)^3-5 \cdot 2^{2m-5} \cdot (2^{m-1}-1)^2-A_5^{\perp} \cdot 2^{2m-6} \cdot 120.\\
\end{split}
\end{equation*}
Solving this equation, we obtain $A_5^{\perp}=(11\cdot2^m+2^{3m-4}-13\cdot 2^{2m-3}-2^4)/120\neq 0$. Hence,
$(\C(f)^{\bar{D}})^\perp$ has parameters $[2^{m-1}-1,2^{m-1}-2m,5]$.

\noindent {\bf Case 2:} $\nu=1$, i.e., $D=\{x \in \mathbb{F}_{2^m}^{*} : {\rm Tr}(\lambda f(x))=1\}$. By (\ref{Nab01400}) and (\ref{eq:pm}), when $(a,b)$ runs over $\mathbb{F}_{2^m}^2\setminus\{(0,0),(\lambda,0)\}$,
the possible values of $\wt(\mathbf{c}(a,b))$ are
$$\frac{n}{2},\,\, \frac{n}{2} \pm 2^{\frac{m-1}{2}}\,\, \text{and}\,\,\frac{n}{2} \pm 2^{\frac{m-3}{2}},$$
where $n$ was given in Lemma \ref{eqadsd}. Moreover, $\wt(\mathbf{c}(a,b))=0$ if and only if $(a,b)=(0,0)$ and
$\wt(\mathbf{c}(a,b))=n$ if $(a,b)=(\lambda,0)$. So, the dimension of $\C(f)^{\bar{D}}$ is $2m$.

Denote
$w_1=2^{m-2}$, $w_2=2^{m-2}+2^{\frac{m-1}{2}}$, $w_3=2^{m-2}-2^{\frac{m-1}{2}}$, $w_4=2^{m-2}+2^{\frac{m-3}{2}}$ and $w_5=2^{m-2}-2^{\frac{m-3}{2}}$. Let $A_{w_i}$ be the number of the codewords with weight $w_i$ in $\C(f)^{\bar{D}}$.
From Lemma \ref{lemdis} we know that $A_1^{\perp}=A_2^{\perp}=A_3^{\perp}=A_4^{\perp}=0$. Then the first five Pless power
moments lead to the following system of equations:
\begin{equation*}
\begin{split}
\begin{cases}
\sum_{i=1}^5A_{w_i}=2^{2m}-2;\\
\sum_{i=1}^5w_iA_{w_i}=2^{2m-1}n-n;\\
\sum_{i=1}^5w_i^2A_{w_i}=2^{2m-2}n(n+1)-n^2;\\
\sum_{i=1}^5w_i^3A_{w_i}=2^{2m-3}n^2(n+3)-n^3;\\
\sum_{i=1}^5w_i^4A_{w_i}=2^{2m-4}n(n+1)(n^2+5n-2)-n^4.
\end{cases}
\end{split}
\end{equation*}
Solving this system of equations, we obtain the desired values of $A_{w_1}$, $A_{w_2}$, $A_{w_3}$, $A_{w_4}$ and $A_{w_5}$  in Table \ref{Table2}.

We now determine the parameters of the dual of $\C(f)^{\bar{D}}$. We treat only the case $n=2^{m-1}$ and the other two cases can be treated similarly. Substituting the value of $n=2^{m-1}$ in Table \ref{Table2}, we obtain that  $A_{w_1}=3\cdot2^{2m-3}+2^{m-2}-2$, $A_{w_2}=A_{w_3}=2^{2m-4}-2^{m-3}$ and $A_{w_4}=A_{w_5}=2^{2m-2}$.
If $d_H\left((\C(f)^{\bar{D}})^\perp  \right)>6$, then
\begin{equation*}
\begin{split}
\sum_{i=0}^{3}\left(
\begin{array}{cccc}
   2^{m-1}  \\
     i  \\
\end{array}
\right)=1+2^{m-1}+2^{m-2}\cdot(2^{m-1}-1)+\frac{2^{m-2}\cdot(2^{m-1}-1)\cdot(2^{m-1}-2)}{3}>2^{2m},
\end{split}
\end{equation*}
which contradicts the sphere packing bound. From Lemma \ref{lemdis}, we then deduce that
$d_H\left( (\C(f)^{\bar{D}})^\perp \right)=6$, and $(\C(f)^{\bar{D}})^\perp$ is distance-optimal with respect to the sphere packing bound.
$\square$

\begin{example}\label{example1}
Let $m=7$ and $f(x)$ be an almost bent function from $\mathbb{F}_{2^7}$ to $\mathbb{F}_{2^7}$ with $W_f(1,0)=2^{\frac{7+1}{2}}$. Let $\C(f)^{\bar{D}}$ be the linear code in Theorem \ref{Theorem1}.
\begin{description}
\item{(1)} If $\nu=0$, then $\C(f)^{\bar{D}}$ has parameters $[71,13,28]$ and its dual has parameters $[71,58,5]$.
\item{(2)} If $\nu=1$  then $\C(f)^{\bar{D}}$ has parameters $[56,14,20]$ and its dual has parameters $[56,42,6]$.
\end{description}
The four codes are optimal according to the tables of best codes known in \cite{Grassl2006}.
\end{example}

\begin{remark}
In \cite{Li2020}, the authors proposed the following open problem (Problem 4.4): Let $\lambda \in \mathbb{F}_{2^s}^*$, $F$ be a function from $\mathbb{F}_{2^m}$ to $\mathbb{F}_{2^s}$ and $D$ be the support of ${\rm Tr}_1^s(\lambda F(x))$. Define a linear code $\C'(F)^{\bar{D}}$ over $\mathbb{F}_2$ by
$$\C'(F)^{\bar{D}}=\{ ({\rm Tr}_1^m(xh)+{\rm Tr}_1^s(yF(h)))_{h \in D}\,:\,x \in \mathbb{F}_{2^m}, y \in \mathbb{F}_{2^s}\}.$$ Determining the weight distributions of the linear codes if $F$ is a vectorial bent function with $m\neq 2s$ or an almost bent function but not the Gold type. Clearly, if $F$ is an almost bent function, then $s=m$. Table \ref{Table2} in Theorem \ref{Theorem1} has given the weight distribution of $\C'(F)^{\bar{D}}$ for $F$ being an almost bent function.
\end{remark}

The following is a list of known almost bent monomials $f(x)=x^d$ on $\mathbb{F}_{2^m}$ for an odd $m$:
\begin{itemize}
\item $d=2^h+1$, where $\gcd(m,h)=1$ is odd \cite{ZR1968};
\item $d=2^{2h}-2^h+1$, where $h\geq 2$ and $\gcd(m,h)=1$ is odd \cite{Kasami2004};
\item $d=2^{\frac{m-1}{2}}+3$, where $m$ is odd\cite{Kasami2004};
\item $d=2^{\frac{m-1}{2}}+2^{\frac{m-1}{4}}-1$, where $m\equiv 1 \pmod 4$ \cite{Hollmann,Hou2004};
\item $d=2^{\frac{m-1}{2}}+2^{\frac{3m-1}{4}}-1$, where $m\equiv 3 \pmod 4$\cite{Hollmann,Hou2004}.
\end{itemize}
All almost bent monomials $f(x)=x^d$ for $d$ in the list above are permutation polynomials on $\mathbb{F}_{2^m}$. Hence, the length of
$\C(f)^{\bar{D}}$ is $n=2^{m-1}-1$ if $\nu=0$ and $n=2^{m-1}$ if $\nu=1$, respectively. Substituting the value of $n$ into  Theorem \ref{Theorem1}, we obtain the following results.

\begin{corollary}\label{coroll}
Let $\C(f)^{\bar{D}}$ be the linear code defined in (\ref{eq:cxxx}) with the position set $D$  in (\ref{eq:cxxdx}). If $f(x)=x^d$ for some integer $d$ in the list above, then the following statements hold.
\begin{description}
\item{(1)} If $\nu=0$, then $\C(f)^{\bar{D}}$ is a $[2^{m-1}-1,2m-1,2^{m-2}-2^{\frac{m-3}{2}}]$ code with the weight distribution in Table~\ref{Table3}. Its dual has parameters $[2^{m-1}-1,2^{m-1}-2m,5]$.
\begin{table}[h]
\caption{\rm Weight distribution of the code $\C(f)^{\bar{D}}$ for $\nu=0$ in Corollary \ref{coroll} }\label{Table3}
\begin{center}
\begin{tabular}{cccc}\hline
     Weight & Multiplicity \\\hline
  $0$ & $1$  \\
 $2^{m-2}$ & $3\cdot2^{2m-4}+2^{m-3}-1$ \\
  $2^{m-2}\pm2^{\frac{m-1}{2}}$ & $2^{2m-5}\mp2^{\frac{3m-7}{2}}\pm2^{\frac{m-5}{2}}-2^{m-4}$ \\
    $2^{m-2}\pm2^{\frac{m-3}{2}}$ & $2^{2m-3}\mp2^{\frac{3m-5}{2}}$ \\
     \hline
\end{tabular}
\end{center}
\end{table}
\item{(2)} If $\nu=1$, then $\C(f)^{\bar{D}}$ is a $[2^{m-1},2m,2^{m-2}-2^{\frac{m-3}{2}}]$ code with the weight distribution in Table~\ref{Table4}. Its dual has parameters $[2^{m-1},2^{m-1}-2m,6]$, and  is distance-optimal
with respect to the sphere packing bound.
  \begin{table}[h]
\caption{\rm Weight distribution of the code $\C(f)^{\bar{D}}$ for $\nu=1$ in Corollary \ref{coroll} }\label{Table4}
\begin{center}
\begin{tabular}{cccc}\hline
     Weight & Multiplicity \\\hline
  $0$ & $1$  \\
 $2^{m-2}$ & $3\cdot2^{2m-3}+2^{m-2}-2$ \\
  $2^{m-2}\pm2^{\frac{m-1}{2}}$ & $2^{2m-4}-2^{m-3}$ \\
    $2^{m-2}\pm2^{\frac{m-3}{2}}$ & $2^{2m-2}$\\
    $2^{m-1}$ & $1$  \\
     \hline
\end{tabular}
\end{center}
\end{table}
\end{description}
\end{corollary}

\begin{example}\label{example2}
Let $\C(f)^{\bar{D}}$ be the linear code in Corollary \ref{coroll}.
\begin{description}
\item{(1)} If $m=7$, $\nu=0$, then $\C(f)^{\bar{D}}$ has parameters $[63,13,24]$ and its dual has parameters $[63,50,5]$.
\item{(2)} If $m=7$, $\nu=1$,  then $\C(f)^{\bar{D}}$ has parameters $[64,14,24]$ and its dual has parameters $[64,50,6]$.
\end{description}
The four codes are optimal according to the tables of best codes known in \cite{Grassl2006}.
\end{example}

\section{Some punctured codes of binary linear codes from quadratic functions}\label{sec-hubei4}

Let $\C(f)^{\bar{D}}$  be the binary punctured code defined in (\ref{eq:cxxx}) with the position set $D$ in (\ref{eqdd}). It is clear that the length of $\C(f)^{\bar{D}}$ is equal to  $2^{m-1}$, as $|D|=|\{x \in \bF_{2^m}^* : {\rm Tr}_1^m(x) =1 \}|=2^{m-1}$. As shown in (\ref{Nab01400}), the Hamming weight of each codeword in this case can be expressed as
\begin{equation}\label{eq:wtquadratic}
\begin{split}
{\rm wt_H}(\mathbf{c}(a,b))&=2^{m-2}-\frac{1}{4}\left(W_f(a,b)-W_f(a,b+1)\right),
\end{split}
\end{equation}
where $W_f(a,b)$ was given in (\ref{Wfab}). In this section, we investigate the weight distribution of the punctured code
$\C(f)^{\bar{D}}$ with the position set $D$ in (\ref{eqdd}), where $f$ is a quadratic
function in the list below and the parameters of its dual.
\begin{itemize}
\item $f(x)= x^{2^k+1}$, \text{where $k$ is an integer with $1  \leq k \leq m-1$};
\item $f(x)= x^{t_1}+x^{t_2}$, \text{where $3\, | \, m$, $m\geq 9$ and $t_1, t_2 \in \{2^{\frac{m}{3}}+1,2^{\frac{2m}{3}}+1,2^{\frac{2m}{3}}+2^{\frac{m}{3}}\}$ with $ t_1\neq t_2$};
\item $f(x)={\rm Tr}_k^m(x^{2^k+1})$, \text{where $m,k$ are positive integers such that $k \, | \, m$}.
\end{itemize}
When $f(x)= x^{2^k+1}$, the parameters and weight distribution of the binary code $\C(f)$ were settled in \cite{Kasami661, Kasami662}. In this section we will investigate the punctured code  $\C(f)^{\bar{D}}$ with a different position set
$D=\{x \in \bF_{2^m}^* : {\rm Tr}_1^m(x) =1 \}$. It is open if the binary code $\C(f)$ was studied in the literature or not
when $f$ is one of the other two quadratic functions in the list above.

\subsection{The case that $f(x)= x^{2^k+1}$}

In this subsection,
we study the punctured code $\C(f)^{\bar{D}}$ in (\ref{eq:cxxx}) and determine its weight distribution,
where $f(x)=x^{2^k+1}$ and $D=\{x \in \bF_{2^m}^* : {\rm Tr}_1^m(x) =1 \}$.
When $k=0$,  $f(x)=x^2$. In this case, it  can be proved that  the punctured code $\C(f)^{\bar{D}}$
is permutation-equivalent to the first-order Reed-Muller code.
In the following, we investigate the linear code $\C(f)^{\bar{D}}$ for $f(x)= x^{2^k+1}$ with $1 \leq k <m$. We start with the following two lemmas.

\begin{lemma}\label{lemd}
Let $\C(f)^{\bar{D}}$ be the linear code defined in (\ref{eq:cxxx}) with the position set $D$ in (\ref{eqdd}). Let $A_i^{\perp}$ denote the number of codewords with weight $i$ in $(\C(f)^{\bar{D}})^\perp$. If $f(x)=x^{2^k+1}$ with $1 \leq k <m$, then
$$A_1^{\perp}=A_2^{\perp}=A_3^{\perp}=A_5^{\perp}=0\,\, \text{and} \,\,A_4^{\perp}=\frac{2^{m-1}\cdot(2^{m-2}-1)\cdot(2^\ell-2)}{4!},$$
where $\ell=\gcd(k,m)$.
\end{lemma}
{\it Proof.} From the definition of the linear code $\C(f)^{\bar{D}}$, we know that $A_i^{\perp}$ is equal to the number of sets  $\{ x_1, x_2,\cdots,x_i\}$ with $i$ pairwise-distinct nonzero elements in $\bF_{2^m}$ such that
\begin{equation*}
\begin{split}
\begin{cases}
{\rm Tr}(x_1)={\rm Tr}(x_2)=\cdots={\rm Tr}(x_i)=1,\\
x_1+x_2+\cdots+x_i=0,\\
 x_1^{2^k+1}+ x_2^{2^k+1}+\cdots+x_i^{2^k+1}=0.\\
\end{cases}
\end{split}
\end{equation*}
It is clear that $A_1^{\perp}=A_2^{\perp}=0$. From the first and second equations, we see that $A_i^\perp=0$ if $i$ is odd.
Hence, $A_3^{\perp}=A_5^{\perp}=0$. In the following, we determine the value of $A_4^{\perp}$, which is equal to the number of sets $\{x_1,x_2,x_3,x_4\}$ with $4$ pairwise-distinct nonzero elements in $\bF_{2^m}$ such that
\begin{equation}\label{eq:numbertupl}
\begin{split}
\begin{cases}
{\rm Tr}(x_1)={\rm Tr}(x_2)={\rm Tr}(x_3)={\rm Tr}(x_4)=1,\\
x_1+x_2+x_3+x_4=0,\\
x_1^{2^k+1}+x_2^{2^k+1}+ x_3^{2^k+1}+x_4^{2^k+1}=0.\\
\end{cases}
\end{split}
\end{equation}
Assume that $x_1=\mu$, $x_2=\mu+\beta$, $x_3=\gamma$ and $x_4=\gamma+\beta$, where $\mu\neq0, \beta, \gamma, \gamma+\beta$, and $\gamma\neq 0, \beta$, and $\beta\neq 0$. From (\ref{eq:numbertupl}) we know that
$A_4^{\perp}$ is equal to the number of the sets of the form $\{\mu,\mu+\beta,\gamma,\gamma+\beta \}$ such that
$$\mu^{2^k+1}+(\mu+\beta)^{2^k+1}=\gamma^{2^k+1}+(\gamma+\beta)^{2^k+1},\,\,  {\rm Tr}(\mu)={\rm Tr}(\gamma)=1\,\,\text{and}\,\, {\rm Tr}(\beta)=0,$$
i.e.,
$$(\mu+\gamma)^{2^{k}-1}=\beta^{2^{k}-1},\,\,{\rm Tr}(\mu)={\rm Tr}(\gamma)=1\,\,\text{and}\,\, {\rm Tr}(\beta)=0.$$
It is clear that $(\mu+\gamma)^{2^{k}-1}=\beta^{2^{k}-1}$ if and only if there is a $\delta \in \mathbb{F}_{2^\ell}$ such that $\mu+\gamma=\delta \beta$ as $\gcd(2^m-1,2^{k}-1)=2^\ell-1$. Then $A_4^{\perp}$ is equal to the number of the sets of the form $\{\mu,\mu+\beta,\mu+ \delta\beta,\mu+\beta(\delta+1) \}$ such that ${\rm Tr}(\mu)=1$ and ${\rm Tr}(\beta)={\rm Tr}(\delta\beta)=0$,
where $\delta\in \mathbb{F}_{2^\ell}\backslash \{0,1\}$, $\mu\neq0,\beta, \delta\beta, \beta(\delta+1)$ and $\beta \neq 0$. Hence,
\begin{equation*}
\begin{split}
A_4^{\perp}&=\frac{1}{8\cdot 4!}\sum_{z_0\in \mathbb{F}_2}\sum_{\mu\in \mathbb{F}_{2^m}^*\backslash \{\beta, \delta\beta, \beta(\delta+1)\}}(-1)^{z_0({\rm Tr}(\mu)-1)}\sum_{z_1\in \mathbb{F}_2}\sum_{\beta\in \mathbb{F}_{2^m}^* }(-1)^{z_1{\rm Tr}(\beta)}\sum_{z_2\in \mathbb{F}_2}\sum_{\delta\in \mathbb{F}_{2^\ell}^*\backslash\{1\}}(-1)^{z_2{\rm Tr}(\delta \beta)}\\
&=\frac{2^{m-3}}{ 4!}\sum_{z_1\in \mathbb{F}_2}\sum_{\beta\in \mathbb{F}_{2^m}^* }(-1)^{z_1{\rm Tr}(\beta)}\sum_{z_2\in \mathbb{F}_2}\sum_{\delta\in \mathbb{F}_{2^\ell}^*\backslash\{1\}}(-1)^{z_2{\rm Tr}(\delta \beta)}\\
&=\frac{2^{m-3}}{ 4!}\sum_{z_1\in \mathbb{F}_2}\sum_{z_2\in \mathbb{F}_2}\sum_{\beta\in \mathbb{F}_{2^m}^* }\sum_{\gamma\in \mathbb{F}_{2^\ell}^*\backslash\{1\}}(-1)^{{\rm Tr}((z_1+z_2\gamma) \beta)}\\
&=\frac{2^{m-3}}{4!}\left(\sum_{z_1\in \mathbb{F}_2}\sum_{z_2\in \mathbb{F}_2}\sum_{\beta\in \mathbb{F}_{2^m} }\sum_{\gamma\in \mathbb{F}_{2^\ell}^*\backslash\{1\}}(-1)^{{\rm Tr}((z_1+z_2\gamma) \beta)}-2^2\cdot(2^\ell-2)\right)\\
&=\frac{2^{m-3}}{4!}\left(2^m\cdot(2^\ell-2)-2^2\cdot(2^\ell-2)\right)=\frac{2^{m-1}\cdot(2^{m-2}-1)\cdot(2^\ell-2)}{4!}.
\end{split}
\end{equation*}
The desired conclusion then follows.  $\square$

\begin{theorem}\label{Theorem2}
Let $\C(f)^{\bar{D}}$ be the linear code defined in (\ref{eq:cxxx}) with the position set $D$ in (\ref{eq:cxxdx}). Let $k$ be a positive integer with $k<m$ and $\ell=\gcd(k,m)$. If $f(x)= x^{2^k+1}$, then the following statements hold.
\begin{description}
\item{(1)} If $v_2(m)\leq v_2(k)$, then $\C(f)^{\bar{D}}$ is a  $[2^{m-1},2m,2^{m-2}-2^{\frac{m+\ell-4}{2}}]$ code with the weight distribution in Table~\ref{Table5}. If $\ell \geq 2$, then its dual has parameters $[2^{m-1},2^{m-1}-2m,4]$. If $\ell=1$, then its dual has parameters $[2^{m-1},2^{m-1}-2m,6]$, and is distance-optimal with respect to the sphere packing bound.
\begin{table}[h]
{\caption{   The weight distribution of $\C(f)^{\bar{D}}$ in Theorem \ref{Theorem2} }\label{Table5}
\begin{center}
\begin{tabular}{cccc}\hline
     Weight & Multiplicity \\\hline
  $0$ & $1$  \\
 $2^{m-2}$ & $2^{2m}-2^{2m-\ell+1}+3\cdot2^{2m-2\ell-1}+2^{m-\ell-1}-2$ \\
  $2^{m-2}\pm2^{\frac{m+\ell-2}{2}}$ & $2^{2m-2\ell-2}-2^{m-\ell-2}$ \\
    $2^{m-2}\pm2^{\frac{m+\ell-4}{2}}$ & $2^{2m-\ell}-2^{2m-2\ell}$\\
    $2^{m-1}$ & $1$  \\
     \hline
\end{tabular}
\end{center}}
\end{table}
\item{(2)} If $v_2(m)> v_2(k)$ and $\gcd(m,k)=1$, then $\C(f)^{\bar{D}}$ is a $[2^{m-1},2m,2^{m-2}-2^{\frac{m}{2}}]$ code with the weight distribution in Table~\ref{Table6}. Its dual has parameters $[2^{m-1}, 2^{m-1}-2m,6]$, and is distance-optimal with respect to the sphere packing bound.
\begin{table}[h]
{\caption{The weight distribution of $\C(f)^{\bar{D}}$ in Theorem \ref{Theorem2}}\label{Table6}
\begin{center}
\begin{tabular}{cccc}\hline
     Weight & Multiplicity \\\hline
  $0$ & $1$  \\
 $2^{m-2}$ & $17\cdot 2^{2m-5}+3\cdot 2^{m-3}-2$ \\
  $2^{m-2}\pm 2^{\frac{m}{2}}$ & $\frac{1}{3}\left( 2^{2m-6}-2^{m-4}\right)$\\
    $2^{m-2}\pm 2^{\frac{m-2}{2}}$ & $\frac{1}{6}\left(11\cdot 2^{2m-3}-2^m\right)$\\
    $2^{m-1}$ & $1$  \\
     \hline
\end{tabular}
\end{center}}
\end{table}
\item{(3)} If $k=\frac{m}{2}$ and $m\geq 4$, then $\C(f)^{\bar{D}}$ is a $[2^{m-1},\frac{3m}{2},2^{m-2}-2^{\frac{m-2}{2}}]$ code with the weight distribution in Table~\ref{Table7}. Its dual has parameters $[2^{m-1}, 2^{m-1}-\frac{3m}{2},4]$,
and is distance-optimal with respect to the sphere packing bound.
\begin{table}[h]
{\caption{The weight distribution of $\C(f)^{\bar{D}}$ in Theorem \ref{Theorem2}}\label{Table7}
\begin{center}
\begin{tabular}{cccc}\hline
     Weight & Multiplicity \\\hline
  $0$ & $1$  \\
 $2^{m-2}$ & $2^{\frac{3m}{2}-1}+2^{m-1}-2$ \\
  $2^{m-2}\pm 2^{\frac{m-2}{2}}$ & $2^{\frac{3m}{2}-2}-2^{m-2}$\\
    $2^{m-1}$ & $1$  \\
     \hline
\end{tabular}
\end{center}}
\end{table}
\end{description}
\end{theorem}
{\it Proof.}  We  prove the desired conclusions for  Cases~(1) and (3) only.
The conclusions in Case (2) can be proved in a similar way.
If $a=0$, it is easy to see that
\[{\rm wt_H}\left( {\bf c}(a,b)\right) = 2^{m-2}-\frac{1}{4}\left( W_f(0,b)- W_f(0,b+1) \right) =  \begin{cases}
 0, &{\rm if }\,\,\, b=0, \\
 2^{m-1}, & {\rm if }\,\,\,  b=1, \\
 2^{m-2}, & {\rm if }\,\,\,  b\neq 0,\,\, 1.
 \end{cases}  \]
If $a\neq0$ and $v_2(m)\leq v_2(k)$, then Lemma \ref{lemeven1} shows that $W_f(a,b) \in \{0, \pm2^{\frac{m+\ell}{2}}\}$.
Consequently, in this case we have $W_f(a,b)- W_f(a,b+1)\in \{0, \pm 2^{\frac{m+\ell}{2}},\pm 2^{\frac{m+\ell+2}{2}}\}$. From (\ref{eq:wtquadratic}) we see that the set of possible nonzero weights of $\C(f)^{\bar{D}}$ is $\{ 2^{m-1}, 2^{m-2}, 2^{m-2}\pm 2^{\frac{m+\ell-2}{2}}, 2^{m-2} \pm 2^{\frac{m+\ell-4}{2}}\}$ and $\C(f)^{\bar{D}}$ has dimension $2m$. Set $w_1=2^{m-1}, w_2= 2^{m-2}, w_3=2^{m-2}+2^{\frac{m+\ell-2}{2}}$, $w_4=2^{m-2}-2^{\frac{m+\ell-2}{2}}$, $w_5=2^{m-2}+2^{\frac{m+\ell-4}{2}}$ and
$w_6=2^{m-2}-2^{\frac{m+\ell-4}{2}}$. It is known that $A_{w_1}= 1$. From Lemma~\ref{lemd} and the first five Pless power
moments we have
\begin{equation}\label{eq:ai}
\left\{ \begin{array}{lll}
\sum_{i=2}^6  A_{w_i}=2^{2m}-2;\\
\sum_{i=2}^6 w_i A_{w_i}=2^{m-1}(2^{2m-1}-1);\\
\sum_{i=2}^6 w_i^2 A_{w_i}=2^{2m-2}( 2^{2m-2} + 2^{m-1} -1);\\
\sum_{i=2}^6 w_i^3 A_{w_i}=2^{3m-3}( 2^{2m-2}+ 3\cdot 2^{m-1}-1);\\
\sum_{i=2}^6 w_i^4 A_{w_i}=2^{3m-5}\left( (2^{m-1}+1)(2^{2m-2}+5\cdot 2^{m-1}-2)-(2^{m-2}-1)(2^\ell-1)\right) -2^{4(m-1)}.
 \end{array}\right.
\end{equation}
Solving the linear equations in (\ref{eq:ai}), we get the desired values of $A_{w_i}$ in Table~\ref{Table5}.
If $\ell > 1$, by Lemma \ref{lemd}, $A_4^\perp >0$. Consequently, the dual distance of the code equals $4$.
If $\ell =1$, by Lemma \ref{lemd}, $A_4^\perp = 0$. Since all weights in $(\C(f)^{\bar{D}})^\perp$ are even,
the minimum distance of $(\C(f)^{\bar{D}})^\perp$ is at least $6$. By the sphere packing bound, the minimum
distance of  $(\C(f)^{\bar{D}})^\perp$ cannot be $8$ or more. Consequently, the minimum distance of
$(\C(f)^{\bar{D}})^\perp$ is equal to $6$. This completes the proof of the conclusions in Case~(1).

Next, we prove the conclusions for Case~(3).
Assume that $k=\frac{m}{2}$ and $f(x) = x^{2^{m/2}+1}$, then
\begin{equation}\label{eq:dadfs2}
\begin{split}
W_f^2(a,b)&=\sum_{x_0 \in \mathbb{F}_{2^m}}(-1)^{{\rm Tr}(ax_0^{2^{m/2}+1} +b x_0 )}\sum_{x \in \mathbb{F}_{2^m}}(-1)^{{\rm Tr}(ax^{2^{m/2}+1} +b x )}\\
&=\sum_{x,y \in \mathbb{F}_{2^m}}(-1)^{{\rm Tr}(a(x+y)^{2^{m/2}+1} +b (x+y)+ax^{2^{m/2}+1} +b x )}\\
&=\sum_{x, y \in \mathbb{F}_{2^m}}(-1)^{{\rm Tr}(a(y^{2^{m/2}+1}+xy^{2^{m/2}}+x^{2^{m/2}}y)+by)}\\
&=\sum_{y \in \mathbb{F}_{2^m}}(-1)^{{\rm Tr}(ay^{2^{m/2}+1} +b y )}\sum_{x \in \mathbb{F}_{2^m}}(-1)^{{\rm Tr}(a(xy^{2^{m/2}}+x^{2^{m/2}}y))}\\
&=2^m\sum_{\begin{small}\begin{array}{c}
y \in \mathbb{F}_{2^m}\\
(a+a^{2^{m/2}})y=0\end{array}\end{small}}(-1)^{{\rm Tr}(ay^{2^{m/2}+1}+b y)}\\
& =\begin{cases}
2^mW_f(a,b), & \text{if $a \in \mathbb{F}_{2^{\frac{m}{2}}}$,} \\
2^m, & \text{otherwise}.
\end{cases}
\end{split}
\end{equation}
If $a\in \mathbb{F}_{2^{\frac{m}{2}}}$, then ${\rm Tr}(ay^{2^{m/2}+1})=0$ and the possible values of $W_f(a,b)$ are as follows:
$$ W_f(a,b)=\sum_{y \in \mathbb{F}_{2^m}}(-1)^{{\rm Tr}(a y^{2^{m/2}+1}+by )}=\sum_{y \in \mathbb{F}_{2^m}}(-1)^{{\rm Tr}(by)}=\begin{cases}
2^m, & \text{if $b=0$,} \\
0, & \text{otherwise}.
\end{cases} $$
Hence,
\begin{equation*}
\begin{split}
W_f(a,b)=\begin{cases}
2^m, & \text{if $a \in \mathbb{F}_{2^{\frac{m}{2}}}$  \text{and}  $b=0$,} \\
0, & \text{if $a \in \mathbb{F}_{2^{\frac{m}{2}}}$  \text{and}  $b\neq 0$,} \\
\pm2^{\frac{m}{2}}, & \text{otherwise.}
\end{cases}
\end{split}
\end{equation*}
When $(a,b)$ runs through $\mathbb{F}_{2^m}^2$, we know that $W_f(a,b)- W_f(a,b+1)\in \{0, \pm 2^{\frac{m+2}{2}}, \pm 2^{m}\}$ and the value $2^m$ occurs $2^{\frac{m}{2}}$ times. Then $\wt(\bc(a,b)) \in \{0, 2^{m-2} \pm 2^{\frac{m-2}{2}}, 2^{m-1}\}$ and $\wt(\bc(a,b))=0$ occurs $2^{\frac{m}{2}}$ times by (\ref{eq:dadfs2}).
So, $\C(f)^{\bar{D}}$ has dimension $\frac{3m}{2}$ and we obtain the weight distribution in Table~\ref{Table7} from the first three Pless power
moments. From the sphere packing bound and Lemma \ref{lemd},  the desired conclusions on $\C(f)^{\bar{D}}$ then follow.
In this case, $\ell  = m/2>1$. It then follows from  Lemma \ref{lemd}, $A_4^\perp >0$. Consequently, the dual distance of the code equals $4$.
$\square$

\begin{example}\label{example5}
Let $\C(f)^{\bar{D}}$ be the linear code in Theorem \ref{Theorem2}.
\begin{description}
\item{(1)} Let $m=5$, $k=1$,  then $\C(f)^{\bar{D}}$ has parameters $[16,10,4]$ and its dual has parameters $[16,6,6]$.
\item{(2)} Let $m=8$, $k=4$,  then $\C(f)^{\bar{D}}$ has parameters $[128,12,56]$ and its dual has parameters $[128,116,4]$.
\end{description}
All the four codes are optimal according to the tables of best codes known  in \cite{Grassl2006}.
\end{example}

\subsection{The case that $f(x)= x^{t_1}+x^{t_2}$}

In this subsection, we investigate the weight distribution of the punctured code $\C(f)^{\bar{D}}$ and the parameters of its dual for $f(x)= x^{t_1}+x^{t_2}$, where $3 \, |\, m$, $m\geq 9$ and $t_1, t_2 \in \{2^{\frac{m}{3}}+1,2^{\frac{2m}{3}}+1,2^{\frac{2m}{3}}+2^{\frac{m}{3}}\}$ with $ t_1\neq t_2$. We first determine all possible Hamming weights in $\C(f)^{\bar{D}}$.

\begin{lemma}\label{eq:l2eme}
Let $\C(f)^{\bar{D}}$ be the linear code defined in (\ref{eq:cxxx}) with the position set $D$ in (\ref{eqdd}). Let $3 \,|\, m$, $m\geq9$ and $f(x)= x^{t_1}+x^{t_2}$, where $t_1, t_2 \in \{2^{\frac{m}{3}}+1,2^{\frac{2m}{3}}+1,2^{\frac{2m}{3}}+2^{\frac{m}{3}}\}$ with $ t_1\neq t_2$.  Then $\C(f)^{\bar{D}}$ is a  $[2^{m-1}, \frac{5m}{3}]$ code  with nonzero weights in the set
$\{ 2^{m-2},\, 2^{m-1},\, 2^{m-2}\pm 2^{\frac{2m}{3}-1}\}$.
\end{lemma}
{\it Proof.}  We prove the conclusions only for the case $t_1=2^{\frac{2m}{3}}+1$ and $t_2=2^{\frac{2m}{3}}+2^{\frac{m}{3}}$.
The conclusions in the other two cases can be similarly proved.
In this case, we have
$$ W_f(a,b)=\sum_{x \in \mathbb{F}_{2^m}}(-1)^{{\rm Tr}(a(x^{2^{2m/3}+1}+x^{2^{2m/3}+2^{m/3}})+bx)}.$$
If $a \in \mathbb{F}_{2^{\frac{m}{3}}}$, then $a+a^{2^{{m/3}}}=0$ and $a+a^{2^{{2m/3}}}=0$. In this case,
$$ W_f(a,b)=\sum_{x \in \mathbb{F}_{2^m}}(-1)^{{\rm Tr}(bx)}=\left\{ \begin{array}{lll}
2^m, \,\,&\text{if $b=0$},\\
0, \,\,&\text{if $b\neq 0$.}
\end{array}\right.$$
Hence, when $(a,b)$ runs over $\mathbb{F}_{2^{\frac{m}{3}}} \times \mathbb{F}_{2^m}$, we obtain
\begin{equation}\label{eqqqqwew}
W_f(a,b)-W_f(a,b+1)=\left\{ \begin{array}{lll}
0, \,\,&\text{occuring $2^{2m}-2^{\frac{m}{3}+1}$ times},\\
2^m, \,\,&\text{occuring $2^{\frac{m}{3}}$ times},\\
-2^m,\,\,&\text{occuring $2^{\frac{m}{3}}$ times}.\\
\end{array}\right.
\end{equation}
If $a \in \mathbb{F}_{2^m}\setminus \mathbb{F}_{2^{\frac{m}{3}}}$, similar to the calculations in (\ref{eq:dadfs2}), we have
\begin{equation*}
\begin{split}
W_f^2(a,b)
&=\sum_{y \in \mathbb{F}_{2^m}}(-1)^{{\rm Tr}(a(y^{2^{2m/3}+1}+y^{2^{2m/3}+2^{m/3}})+by)}\sum_{x \in  \mathbb{F}_{2^m}}(-1)^{{\rm Tr}((ay^{2^{m/3}}+a^{2^{m/3}}y+a^{2^{2m/3}}y^{2^{m/3}}+ay)x^{2^{2m/3}})}\\
&=2^m\sum_{\begin{small}\begin{array}{c}
y \in \mathbb{F}_{2^m}\\
(a+a^{2^{{2m/3}}})y^{2^{m/3}}+(a^{2^{m/3}}+a)y=0\end{array}\end{small}}(-1)^{{\rm Tr}(a(y^{2^{2m/3}+1}+y^{2^{2m/3}+2^{m/3}})+by)}.
\end{split}
\end{equation*}
Let $L_a(y)=(a+a^{2^{2m/3}})y^{2^{m/3}}+(a^{2^{m/3}}+a)y$, then
$$ \textrm{Ker} (L_a(y))=\{ \, y \in \mathbb{F}_{2^m} \,\, | \,\, L_a(y)=0 \, \}=\left\{ (a^{2^{2m/3}}+a)z:\,\, z \in \mathbb{F}_{2^{\frac{m}{3}}}\right\}.$$
From (\ref{eqWf2}) we get
$$ W_f^2(a,b)=\begin{cases}
2^{\frac{4m}{3}}, & \text{if ${{\rm Tr}\left(a(y^{2^{2m/3}+1}+y^{2^{2m/3}+2^{m/3}}) +by\right)}=0$ for all $y\in$Ker$(L_a(y))$,} \\
0, & \text{otherwise}.
\end{cases} $$
If ${{\rm Tr}\left(a(y^{2^{2m/3}+1}+y^{2^{2m/3}+2^{m/3}}) +by\right)}=0$ for all $y\in$Ker$(L_a(y))$, then $${{\rm Tr}\left(a(y^{2^{2m/3}+1}+y^{2^{2m/3}+2^{m/3}}) +(b+1)y\right)}={\rm Tr}(y)={\rm Tr}_1^{\frac{m}{3}}\left({\rm Tr}_{\frac{m}{3}}^m((a^{2^{2m/3}}+a)t)\right)=0$$ because $t \in \mathbb{F}_{2^{\frac{m}{3}}}$.
Hence,
$W_f(a,b)- W_f(a,b+1) \in \left\{0, \pm 2^{\frac{2m}{3}+1}\right\}$ for $a \in \mathbb{F}_{2^m}\backslash \mathbb{F}_{2^{\frac{m}{3}}}$.  Combining this with (\ref{eqqqqwew}), when $(a,b)$ runs through $\mathbb{F}_{2^m} \times \mathbb{F}_{2^m}$, we have
 $$W_f(a,b)- W_f(a,b+1)\in\left\{0, \pm2^m, \pm 2^{\frac{2m}{3}+1}\right\}$$
and each of the values $\pm2^m$ occurs $2^{\frac{m}{3}}$ times. Then from (\ref{eq:wtquadratic}) we know that
${\rm wt_H}(\bc(a,b))=0$ and ${\rm wt_H}(\bc(a,b))=2^{m-1}$ both occur $2^{\frac{m}{3}}$ times and the nonzero weights
in $\C(f)^{\bar{D}}$ belong to the set $\{ 2^{m-2},\, 2^{m-1},\, 2^{m-2}\pm 2^{\frac{2m}{3}-1}\}$. It then follows that
 $\C(f)^{\bar{D}}$ is degenerate and has dimension $\frac{5m}{3}$. This completes the proof.  $\square$

\begin{theorem}\label{Theorem3}
Follow the notation and conditions introduced in Lemma \ref{eq:l2eme}.
Then $\C(f)^{\bar{D}}$ is a $[2^{m-1},\frac{5m}{3},2^{m-2}-2^{\frac{2m}{3}-1}]$ code with the weight distribution in
Table~\ref{Table8}. Its dual has parameters $[2^{m-1},2^{m-1}-\frac{5m}{3},4]$, and  is distance-optimal with respect to the sphere packing bound.
\begin{table}[ht]
{\caption{\rm   The weight distribution of $\C(f)^{\bar{D}}$ in Theorem \ref{Theorem3} }\label{Table8}
\begin{center}
\begin{tabular}{cccc}\hline
     Weight & Multiplicity \\\hline
 $0$ & $1$  \\
 $2^{m-2}$ & $2^{\frac{5m}{3}}-2^{\frac{4m}{3}-1}+2^{\frac{2m}{3}-1}-2$ \\
  $2^{m-2}\pm 2^{\frac{2m}{3}-1}$ & $2^{\frac{4m}{3}-2}-2^{\frac{2m}{3}-2}$ \\
  $2^{m-1}$ & $1$\\
     \hline
\end{tabular}
\end{center}}
\end{table}
\end{theorem}
{\it Proof.}   From Lemma \ref{eq:l2eme}, we conclude that the dimension of $\C(f)^{\bar{D}}$ is $\frac{5m}{3}$, the possible weights in $\C(f)^{\bar{D}}$ are given in the set $\{0, 2^{m-1}, 2^{m-2}, $ $2^{m-2} \pm 2^{\frac{2m}{3}-1}\}$ and the weight $2^{m-1}$ occurs $1$ time.

Denote $w_1=2^{m-2}$, $w_2=2^{m-2}-2^{\frac{2m}{3}-1}$ and $w_3=2^{m-2}+2^{\frac{2m}{3}-1}$. Let $A_{w_i}$ be the number of the codewords with weight $w_i$ in $\C(f)^{\bar{D}}$.
Note that the all-one vector is a codeword of  $\C(f)^{\bar{D}}$. It then follows that all codewords in  $(\C(f)^{\bar{D}})^\perp$ have even weights. It is easily seen that the minimum weight in $(\C(f)^{\bar{D}})^\perp$ cannot be $2$. Consequently, the minimum weight
in $(\C(f)^{\bar{D}})^\perp$ is at least $4$. From the first three Pless power moments, we have
\begin{equation*}
\begin{split}
\begin{cases}
\sum_{i=1}^3A_{w_i}=2^{\frac{5m}{3}}-2;\\
\sum_{i=1}^3w_iA_{w_i}=2^{\frac{8m}{3}-2}-2^{m-1};\\
\sum_{i=1}^3w_i^2A_{w_i}=2^{\frac{8m}{3}-2}(2^{m-1}+1)-2^{2m-2}.
\end{cases}
\end{split}
\end{equation*}
Solving this system of equations, we obtain $A_{w_1}=2^{\frac{5m}{3}}-2^{\frac{4m}{3}-1}+2^{\frac{2m}{3}-1}-2$, $A_{w_2}=A_{w_3}=2^{\frac{4m}{3}-2}-2^{\frac{2m}{3}-2}$.

We now consider the minimum distance of $(\C(f)^{\bar{D}})^\perp$. We have already proved that
$$
d_H\left( (\C(f)^{\bar{D}})^\perp\right) \geq 4.
$$
 If there exists a $[2^{m-1},2^{m-1}-\frac{5m}{3}]$ binary code with Hamming distance at least $5$, then
\begin{equation*}
\begin{split}
\sum_{i=0}^{2}\left(
\begin{array}{cccc}
   2^{m-1}  \\
     i  \\
\end{array}
\right)=1+2^{m-1}+2^{m-2}\cdot(2^{m-1}-1)>2^{\frac{5m}{3}},
\end{split}
\end{equation*}
 which contradicts the sphere packing bound.  Hence, $d_H\left( (\C(f)^{\bar{D}})^\perp \right)=4$ and $(\C(f)^{\bar{D}})^\perp$
  is distance-optimal to the sphere packing bound.
  $\square$

\begin{example}\label{example4}
Let $\C(f)^{\bar{D}}$ be the linear code in Theorem \ref{Theorem3}.
 Let $m=9$, then $\C(f)^{\bar{D}}$ has parameters $[256,15,96]$ and its dual has parameters $[256,241,4]$.
\end{example}

We settled the weight distribution of the punctured code $\C(f)^{\bar{D}}$ in Theorem \ref{Theorem3}, but do not know
if the corresponding code $\C(f)$ was studied in the literature or not.

\vskip 6pt
\subsection{The case that $f(x)={\rm Tr}_k^m(x^{2^k+1})$}
\vskip 6pt

In this subsection, we study the weight distribution of the punctured code $\C(f)^{\bar{D}}$ and the parameters of
its dual for $f(x)={\rm Tr}_k^m(x^{2^k+1})$, where $k$ divides $m$. It is easy to see that $f(x)=0$ if $k=\frac{m}{2}$.
In the following, we just consider the case that $k\not\in \{m, \frac{m}{2}\}$. We begin with the following lemma.

\begin{lemma}\label{eq:lem}
Let $\C(f)^{\bar{D}}$ be the punctured code defined in (\ref{eq:cxxx}) with the position set $D$ in (\ref{eqdd}). Let $f(x)={\rm Tr}_k^m(x^{2^k+1})$, where $k$ divides $m$ and $k\not\in \{m, \frac{m}{2}\}$.  Let $t=2^{\frac{m+2k-2}{2}}$ if  $v_2(m)> v_2(k)+1$, and $t=2^{\frac{m+2k-4}{2}}$ if  $v_2(m)= v_2(k)+1$, and $t=2^{\frac{m+k-4}{2}}$ if  $v_2(m)= v_2(k)$.  Then
  $\C(f)^{\bar{D}}$ is a  $[2^{m-1}, k+m]$ code whose nonzero weights are in the set $\left\{ 2^{m-2},\,\, 2^{m-1},\,\, 2^{m-2}\pm t\right\}$.
\end{lemma}
{\it Proof.} We prove the conclusions only for the case $v_2(m)>v_2(k)+1$. The conclusions for the other two cases can be
similarly proved. We first determine the possible values of $W_f(a,b)$ for $(a,b)\in \mathbb{F}_{2^m}^2$, where $W_f(a,b)$ was defined in (\ref{Wfab}). Note that ${\rm Tr}(a {\rm Tr}_k^m(x^{2^k+1}))={\rm Tr}({\rm Tr}_k^m(a)x^{2^k+1})$. If ${\rm Tr}_k^m(a)=0$,  then
$$W_f(a,b)=\sum_{x \in \mathbb{F}_{2^m}}(-1)^{{\rm Tr}(bx)}=\left\{ \begin{array}{lll}
2^m, \,\,&\text{if $b=0$},\\
0, \,\,&\text{if $b\neq 0$.}
\end{array}\right.$$
Let $L=\{a \in \mathbb{F}_{2^m} : {\rm Tr}_k^m(a)=0\}$, then $|L|=2^{m-k}$. Hence, when $(a,b)$ runs over $L \times \mathbb{F}_{2^m}$, we have
\begin{equation}\label{eq:wt01}
W_f(a,b)- W_f(a,b+1)=\left\{ \begin{array}{lll}
0, \,\,&\text{occuring $2^{m+k}-2^{m-k+1}$ times},\\
2^m, \,\,&\text{occuring $2^{m-k}$ times},\\
-2^m,\,\,&\text{occuring $2^{m-k}$ times}.\\
\end{array}\right.
\end{equation}
If ${\rm Tr}_k^m(a)\neq0$, similar to the discussions in (\ref{eq:dadfs2}), we have
\begin{equation*}
\begin{split}
W_f^2(a,b)&=\sum_{x \in \mathbb{F}_{2^m}}(-1)^{{\rm Tr}(a{\rm Tr}_k^m(x^{2^k+1}) +b x)}\sum_{y \in \mathbb{F}_{2^m}}(-1)^{{\rm Tr}(a{\rm Tr}_k^m(xy^{2^k}+x^{2^k}y))}\\
&=2^m\sum_{\begin{small}\begin{array}{c}
x \in \mathbb{F}_{2^m}\\
x+x^{2k}=0\end{array}\end{small}}(-1)^{{\rm Tr}(a{\rm Tr}_k^m(x^{2^k+1}) +b x )}\\
&=2^m \cdot\sum_{x \in \mathbb{F}_{2^{2k}}}(-1)^{{\rm Tr}(a{\rm Tr}_k^m(x^{2^k+1}) +b x )},
\end{split}
\end{equation*}
as $v_2(m)>v_2(k)+1$.
Then by (\ref{eqWf2}) we obtain
$$ W_f^2(a,b)=\begin{cases}
2^{m+2k}, & \text{if ${{\rm Tr}\left(a{\rm Tr}_k^m(x^{2^k+1}) +b x \right)}=0$ for all $x \in \mathbb{F}_{2^{2k}}$,} \\
0, & \text{otherwise}.
\end{cases} $$
Clearly, if ${{\rm Tr}(a{\rm Tr}_k^m(x^{2^k+1}) +bx)}=0$ for all $x \in \mathbb{F}_{2^{2k}}$, then $ {{\rm Tr}(a{\rm Tr}_k^m(x^{2^k+1}) +(b+1)x)}={{\rm Tr}\left(x\right)}=\frac{m}{2k}{{\rm Tr}_1^{2k}\left(x\right)}$. Hence,
$W_f(a,b)- W_f(a,b+1)\in\{0, \pm 2^{\frac{m+2k+2}{2}}\}$ for ${\rm Tr}_k^m(a)\neq0$. Combining this with (\ref{eq:wt01}), when $(a,b)$ runs through $\mathbb{F}_{2^m}^2$, we have
 $$W_f(a,b)- W_f(a,b+1)\in\left\{0, 2^m, -2^m, \pm 2^{\frac{m+2k+2}{2}}\right\}$$
and  each of the values $\pm2^m$ occurs $2^{m-k}$ times. Then ${\rm wt_H}(\bc(a,b))=0$ and ${\rm wt_H}(\bc(a,b))=2^{m-1}$ both occur $2^{m-k}$ times and every nonzero weight
in $\C(f)^{\bar{D}}$ belongs to the set $\{ 2^{m-2},\, 2^{m-1},\, 2^{m-2}\pm 2^{\frac{m+2k-2}{2}}\}$ by (\ref{eq:wtquadratic}) . Hence, $\C(f)^{\bar{D}}$ is degenerate and has dimension $m+k$.  This completes the proof.  $\square$

Using Lemma \ref{eq:lem} and similar discussions in the proof of Theorem \ref{Theorem2},  one can prove the following
theorem.

\begin{theorem}\label{Theorem4}
Follow the notation and conditions introduced in Lemma \ref{eq:lem}.
 Then $\C(f)^{\bar{D}}$ is a $[2^{m-1},m+k,2^{m-2}-t]$ code with the weight distribution in Table~\ref{Table9}. Its dual has parameters $\left[2^{m-1},2^{m-1}-m-k,4\right]$, and  is distance-optimal with respect to the sphere packing bound.
\begin{table}[ht]
{\caption{\rm   The weight distribution of $\C(f)^{\bar{D}}$ in Theorem \ref{Theorem4} }\label{Table9}
\begin{center}
\begin{tabular}{cccc}\hline
     Weight & Multiplicity \\\hline
  $0$ & $1$  \\
 $2^{m-2}$ & $2^{m+k}-2+2^{2m-3}\cdot (1-2^k)/t^2$ \\
  $2^{m-2}\pm t$ & $2^{2m-4}\cdot (2^k-1)/t^2$ \\
  $2^{m-1}$ & $1$\\
     \hline
\end{tabular}
\end{center}}
\end{table}
\end{theorem}

\begin{example}\label{example3}
Let $\C(f)^{\bar{D}}$ be the linear code in Theorem \ref{Theorem4}. Let $m=5$ and $k=1$. Then $\C(f)^{\bar{D}}$ has parameters $[16,6,6]$ and its dual has parameters $[16,10,4]$. Both codes are optimal according  to the tables of best codes known in \cite{Grassl2006}.
\end{example}

We settled the parameters and weight distribution of the punctured code  $\C(f)^{\bar{D}}$ in
Theorem~\ref{Theorem4}, but do not know if the corresponding code $\C(f)$ was studied in the literature or not.

\section{Some punctured codes of binary linear codes from cyclotomic classes}\label{sec-hubei5}

In this section, we settle the weight distribution of the punctured code $\C(f)^{\bar{D}}$ and the parameters of its dual,
where the position set $D$
is a cyclotomic class and $f(x)=x^d$ for some integer $d$.
Let $\gamma$ be a primitive element of $\mathbb{F}_{2^m}$ and let $t$ be a positive integer dividing $2^m-1$.
Let $D=\langle \gamma^t\rangle$, which is the subgroup of $\mathbb{F}_{2^m}^*$ generated by $\gamma^t$.
The multiplicative cosets of $D$ are called the cyclotomic classes of order $t$ in $\mathbb{F}_{2^m}^*$.
Recall that the  binary punctured code is
\begin{equation}\label{eq:gencdode}
\C(f)^{\bar{D}} = \{ \mathbf{c}(a,b)=({\rm Tr}(ax^d+bx))_{x \in D} :   a, b\in \bF_{2^m}\}
\end{equation}
if $f(x)=x^d$.
Since $|D|=\frac{2^m-1}{t}$, the length $n$ of $\C(f)^{\bar{D}}$ is $\frac{2^m-1}{t}$. It is easily seen that the Hamming weight
of the codeword $\bc(a,b)$ is given by
\begin{equation}\label{eq:wt12}
\begin{split}
{\rm wt_H}( {\bf c}(a,b))&=n-\left|\left\{x\in D:\,\,  {\rm Tr}\left(ax^d+bx\right)=0 \right\}\right|=\frac{1}{2}\left(n-\sum_{x \in D}(-1)^{ {\rm Tr}(ax^d+bx)}\right).
\end{split}
\end{equation}
To determine the weight distribution of $\C(f)^{\bar{D}}$, we need to determine the value distribution of
\begin{equation}\label{eq:genc1dode}
T(a,b)=\sum_{x \in D}(-1)^{ {\rm Tr}(ax^{d}+bx)}
\end{equation}
for $(a,b)$ running through $\mathbb{F}_{2^m}^2$. In the following, we propose several classes of linear codes with few weights by choosing proper $d$ and $t$.

\subsection{The case that  $d=\frac{2^m-1}{3}$ and lcm$(3,t) \, | \, (2^{\frac{m}{2}}+1)$ }

In this subsection, we always assume that $v_2(m)=1$, $d=\frac{2^m-1}{3}$ and $t$ is a positive integer satisfying lcm$(3,t) \, | \,( 2^{\frac{m}{2}}+1)$.
If $3 \, | \, t$, then $x^{\frac{2^m-1}{3}}=1$ for any $x \in D$. From (\ref{eq:genc1dode}) we have
\begin{equation}\label{Nabfwt}
\begin{split}
T(a,b)=\sum_{x \in  D}(-1)^{ {\rm Tr}(a+bx)}.
\end{split}
\end{equation}
If $3 \, \nmid \, t$, then
\begin{equation}\label{Nab000}
\begin{split}
T(a,b)&=\sum_{x \in \langle\gamma^{3t}\rangle}(-1)^{ {\rm Tr}(a+bx)}+\sum_{x \in \langle\gamma^{3t}\rangle}(-1)^{ {\rm Tr}(a\gamma^{\frac{t(2^m-1)}{3}}+b\gamma^tx)}+\sum_{x \in \langle\gamma^{3t}\rangle}(-1)^{ {\rm Tr}(a\gamma^{\frac{2t(2^m-1)}{3}}+b\gamma^{2t}x)}\\
&=\frac{1}{3t}\left(\sum_{x \in\mathbb{F}_{2^m}^*}(-1)^{ {\rm Tr}(a+bx^{3t})}+\sum_{x \in\mathbb{F}_{2^m}^*}(-1)^{ {\rm Tr}(a\gamma^{\frac{t(2^m-1)}{3}}+b\gamma^{t}x^{3t})}+\sum_{x \in\mathbb{F}_{2^m}^*}(-1)^{ {\rm Tr}(a\gamma^{\frac{2t(2^m-1)}{3}}+b\gamma^{2t}x^{3t})}\right)\\
&=\frac{1}{3t}\left(\sum_{x \in\mathbb{F}_{2^m}}(-1)^{ {\rm Tr}(a+bx^{3t})}+\sum_{x \in\mathbb{F}_{2^m}}(-1)^{ {\rm Tr}(a\gamma^{\frac{t(2^m-1)}{3}}+b\gamma^{t}x^{3t})}+\sum_{x \in\mathbb{F}_{2^m}}(-1)^{ {\rm Tr}(a\gamma^{\frac{2t(2^m-1)}{3}}+b\gamma^{2t}x^{3t})}\right)\\
&-\frac{1}{3t}\left((-1)^{{\rm Tr}(a)}+(-1)^{{\rm Tr}(a\gamma^{\frac{t(2^m-1)}{3}})}+(-1)^{{\rm Tr}(a\gamma^{\frac{2t(2^m-1)}{3}})}\right).
\end{split}
\end{equation}
In order to obtain the possible values of  $T(a,b)$ for $3 \, \nmid \, t$, we need the following lemma.

\begin{lemma}\label{eqdad}
Let $N$ be the number of zeros in the sequence $\left({\rm Tr}(a),\,\, {\rm Tr}(a\gamma^{\frac{t(2^m-1)}{3}}), \,\,{\rm Tr}(a\gamma^{\frac{2t(2^m-1)}{3}})\right)$. When $a$ runs over $\mathbb{F}_{2^m}$, we have
\[N=\left\{ \begin{array}{lll}
           3, & {\rm occuring} \,\, \,2^{m-2}\,\,\, {\rm times},\\
         1, & {\rm occuring} \,\, \,3\cdot2^{m-2}\,\, \,{\rm times}. \end{array}  \right.\]
\end{lemma}
{\it Proof.} Obviously, the possible values of $N$ are 0, 1, 2 or 3. Let $N_i$ denote the number of times that $N=i$ when  $a$ runs over $\mathbb{F}_{2^m}$, where $i \in \{0,1,2,3\}$. Then
\begin{equation*}
\begin{split}
N_3&=\frac{1}{2^3}\sum_{a \in \mathbb{F}_{2^m}}\sum_{y_0 \in \mathbb{F}_2}(-1)^{{\rm Tr}(y_1 a)}\sum_{y_1 \in \mathbb{F}_2}(-1)^{{\rm Tr}(y_1 a\gamma^{\frac{2^m-1}{3}})}\sum_{y_2 \in \mathbb{F}_2}(-1)^{{\rm Tr}(y_2 a\gamma^{\frac{2(2^m-1)}{3}})}\\
&=\frac{1}{2^3}\sum_{a \in \mathbb{F}_{2^m}}\sum_{y_0 \in \mathbb{F}_2}\sum_{y_1 \in \mathbb{F}_2}\sum_{y_2 \in \mathbb{F}_2}(-1)^{{\rm Tr}\left( a (y_0+y_1 \gamma^{\frac{2^m-1}{3}}+y_2 \gamma^{\frac{2(2^m-1)}{3}})\right)}.
\end{split}
\end{equation*}
Note that $y_0+y_1 \gamma^{\frac{2^m-1}{3}}+y_2 \gamma^{\frac{2(2^m-1)}{3}}=0$ if and only if $y_0=y_1=y_2=0$ or $y_0=y_1=y_2=1$. Then
$$N_3=\frac{1}{2^3}\left(2^m+2^m\right)=2^{m-2}.$$
Due to symmetry, we have
\begin{eqnarray*}
N_2&=& \frac{3}{2^3}\sum_{a \in \mathbb{F}_{2^m}}\sum_{y_0 \in \mathbb{F}_2}(-1)^{y_0({\rm Tr}( a)-1)}\sum_{y_1 \in \mathbb{F}_2}(-1)^{{\rm Tr}(y_1 a\gamma^{\frac{2^m-1}{3}})}\sum_{y_2 \in \mathbb{F}_2}(-1)^{{\rm Tr}(y_2 a\gamma^{\frac{2(2^m-1)}{3}})}\\
&=& \frac{3}{2^3}\sum_{a \in \mathbb{F}_{2^m}}\sum_{y_1 \in \mathbb{F}_2}\sum_{y_0 \in \mathbb{F}_2}\sum_{y_2 \in \mathbb{F}_2}(-1)^{{\rm Tr}( a (y_0+y_1 \gamma^{\frac{2^m-1}{3}}+y_2 \gamma^{\frac{2(2^m-1)}{3}})-y_0)}\\
&=& \frac{3}{2^3}\sum_{a \in \mathbb{F}_{2^m}}\sum_{y_1 \in \mathbb{F}_2}\sum_{y_2 \in \mathbb{F}_2}(-1)^{{\rm Tr}\left( a (y_1 \gamma^{\frac{2^m-1}{3}}+y_2 \gamma^{\frac{2(2^m-1)}{3}})\right)}- \\
& & \frac{3}{2^3}\sum_{a \in \mathbb{F}_{2^m}}\sum_{y_1 \in \mathbb{F}_2}\sum_{y_2 \in \mathbb{F}_2}(-1)^{{\rm Tr}\left( a (1+y_1 \gamma^{\frac{2^m-1}{3}}+y_2 \gamma^{\frac{2(2^m-1)}{3}})\right)}.
\end{eqnarray*}
Note that  $y_1 \gamma^{\frac{2^m-1}{3}}+y_2 \gamma^{\frac{2(2^m-1)}{3}}=0$ if and only if $y_1=y_2=0$, and $1+y_1 \gamma^{\frac{2^m-1}{3}}+y_2 \gamma^{\frac{2(2^m-1)}{3}}=0$ if and only if $y_1=y_2=1$. Then
$$N_2=\frac{1}{2^3}\left(2^m-2^m\right)=0.$$
Similarly, we can prove that $N_1=3 \cdot 2^{m-2}$ and $N_0=0$. $\square$

\begin{theorem}\label{Theorem5}
Let $v_2(m)=1$, $d=\frac{2^m-1}{3}$ and $t$ be a positive integer satisfying $\textrm{lcm}(3,t) \, | \, (2^{\frac{m}{2}}+1)$.
Let $\C(f)^{\bar{D}}$ be the linear code defined in (\ref{eq:gencdode}) and $D=\langle\gamma^t\rangle$. If $t\neq 2^{\frac{m}{2}}+1$, then the following statements hold.
\begin{description}
\item{(1)} If $3\, | \, t$, then $\C(f)^{\bar{D}}$ is a $[\frac{2^m-1}{t},m+1]$ code with the weight distribution in  Table~\ref{Table10}. Its dual has parameters $[\frac{2^m-1}{t},\frac{2^m-1}{t}-m-1,4]$, and is distance-optimal with respect to the sphere packing bound.

\begin{table}[h]
\caption{\rm Weight distribution of the code $\C(f)^{\bar{D}}$ for  $3\, | \, t$ in Theorem \ref{Theorem5}  }\label{Table10}
\begin{center}
\begin{tabular}{cccc}\hline
Weight      & Multiplicity \\\hline
$0$ &  $1$ \\   \hline
 $\frac{1}{2t}\left( 2^m-2-2^{\frac{m}{2}}\right)$ & $\frac{(t-1)(2^m-1)}{t}$  \\   \hline
 $\frac{1}{2t}\left( 2^m+2^{\frac{m}{2}}\right)$ & $\frac{(t-1)(2^m-1)}{t}$  \\   \hline
  $\frac{1}{2t}\left( 2^m-2+(t-1)2^{\frac{m}{2}}\right)$ & $\frac{(2^m-1)}{t}$ \\   \hline
  $\frac{1}{2t}\left( 2^m-(t-1)2^{\frac{m}{2}}\right)$ & $\frac{(2^m-1)}{t}$ \\   \hline
  $\frac{2^{m}-1}{t}$ & $1$  \\   \hline
\end{tabular}
\end{center}
\end{table}

\item{(2)} If $3\, \nmid \, t$, then $\C(f)^{\bar{D}}$ is a $[\frac{2^m-1}{t},m+2]$ code with the weight distribution in
Table~\ref{Table11}. Its dual has parameters $[\frac{2^m-1}{t},\frac{2^m-1}{t}-m-2,3]$.
\begin{table}[h]
\caption{\rm Weight distribution of the code $\C(f)^{\bar{D}}$ for  $3\, \nmid \, t$ in Theorem \ref{Theorem5} }\label{Table11}
\begin{center}
\begin{tabular}{cccc}\hline
Weight      & Multiplicity \\\hline
$0$ &  $1$ \\   \hline
  $\frac{2^{m}-1}{2t}-\frac{1}{2t}\left((t-1)2^{\frac{m}{2}}-1\right)$ & $\frac{2^m-1}{t}$  \\   \hline
 $\frac{2^{m}-1}{2t}+\frac{1}{6t}\left((3t-1)2^{\frac{m}{2}}-1\right)$ & $\frac{2^{m+1}-2}{t}$  \\   \hline
 $\frac{1}{2t}\left(2^m+2^{\frac{m}{2}}\right)$ & $\frac{(t-1)(2^m-1)}{t}$  \\   \hline
  $\frac{2^{m}-1}{2t}-\frac{1}{6t}\left(2^{\frac{m}{2}}+1\right)$ & $\frac{3(t-1)(2^m-1)}{t}$  \\   \hline
  $\frac{2^{m}-1}{2t}-\frac{1}{6t}\left((3t+1)2^{\frac{m}{2}}+1\right)$ & $\frac{2^m-1}{t}$  \\   \hline
  $\frac{2(2^m-1)}{3t}$ & $3$   \\  \hline
\end{tabular}
\end{center}
\end{table}
\end{description}
\end{theorem}

{\it Proof.} We prove this theorem case by case as follows.

\noindent {\bf Case 1:} $3\, | \, t$. From (\ref{Nabfwt}) we have
\begin{equation*}
\begin{split}
T(a,b)&=(-1)^{{\rm Tr}(a)}\sum_{x \in E}(-1)^{ {\rm Tr}(bx)}=\frac{1}{t}(-1)^{{\rm Tr}(a)}\sum_{x \in \mathbb{F}_{2^m}^*}(-1)^{ {\rm Tr}(bx^{t})}\\
&=\frac{1}{t}(-1)^{{\rm Tr}(a)}-\frac{1}{t}(-1)^{{\rm Tr}(a)}\sum_{x \in \mathbb{F}_{2^m}}(-1)^{ {\rm Tr}(bx^{t})}.
\end{split}
\end{equation*}
If $b=0$, it is clear that
\begin{equation}\label{eq:dasdsad}
\begin{split}
T(a,0)&=\left\{ \begin{array}{lcl}
           \frac{(1-2^m)}{t},  & {\rm if} \,\,\,  {\rm Tr}(a)=0,\\
           \frac{(2^m-1)}{t},
             &  {\rm if} \,\,\,  {\rm Tr}(a)=1. \end{array}  \right.
\end{split}
\end{equation}
If $b\neq 0$,
then $b$ can be written as $b=\gamma^i$, where $\gamma$ is a primitive element of $\mathbb{F}_{2^m}$ and $1\leq i \leq 2^m-1$. From Lemma \ref{lemeven3} we have
\begin{equation}\label{eq:Tdasdsad}
\begin{split}
T(a,\gamma^i)&=\left\{ \begin{array}{lcl}
          \frac{1}{t}(-1)^{{\rm Tr}(a)}-\frac{1}{t}(-1)^{{\rm Tr}(a)}(-1)^{s}2^{\frac{m}{2}}, & {\rm if} \,\,\,  i \not\equiv 0 \pmod t,\\
          \frac{1}{t}(-1)^{{\rm Tr}(a)}-\frac{1}{t}(-1)^{{\rm Tr}(a)}(-1)^{s-1}(t-1)2^{\frac{m}{2}},
             & {\rm if} \,\,\,  i \equiv 0 \pmod t, \end{array}  \right.
\end{split}
\end{equation}
as $t$ is a positive integer such that lcm$(3,t)\, |\, (2^{m/2}+1)$.
Hence, when $(a,b)$ runs over $\mathbb{F}_{2^m}^2$, by (\ref{eq:dasdsad}) and (\ref{eq:Tdasdsad}), the value distribution of $T(a,b)$ is given as follows:
\begin{equation}\label{eq:Tdasdsad1}
\begin{split}
T(a,b)&=\left\{ \begin{array}{lll}
 \frac{(1-2^m)}{t}, & {\rm occuring} \,\,  2^{m-1}\,\, {\rm times},\\
\frac{(2^m-1)}{t},&{\rm occuring} \,\,  2^{m-1}\,\, {\rm times},\\
\frac{1}{t}\left(2^{\frac{m}{2}}+1\right),   &{\rm occuring} \,\, \frac{ 2^{m-1}(t-1)(2^m-1)}{t}\,\, {\rm times},\\
-\frac{1}{t}\left(2^{\frac{m}{2}}+1\right),   &{\rm occuring} \,\, \frac{ 2^{m-1}(t-1)(2^m-1)}{t}\,\, {\rm times},\\
\frac{1}{t}-\frac{1}{t}(t-1)2^{\frac{m}{2}},  &{\rm occuring} \,\, \frac{ 2^{m-1}(2^m-1)}{t}\,\, {\rm times},\\
-\frac{1}{t}+\frac{1}{t}(t-1)2^{\frac{m}{2}},  &{\rm occuring} \,\, \frac{ 2^{m-1}(2^m-1)}{t}\,\, {\rm times}.
                         \end{array}  \right.
\end{split}
\end{equation}
From (\ref{eq:wt12}) and (\ref{eq:Tdasdsad1}), we know that the Hamming weight $0$ occurs $2^{m-1}$ times when $(a,b)$ runs through $\mathbb{F}_{2^m}^2$. Hence, in this case, $\C(f)^{\bar{D}}$ is degenerate and has dimension $m+1$.  Dividing each frequency by $2^{m-1}$ in (\ref{eq:Tdasdsad1}), we get the weight distribution in Table \ref{Table10} from (\ref{eq:wt12}). From the first five Pless power
moments and the weight distribution of $\C(f)^{\bar{D}}$,  we deduce that the dual of $\C(f)^{\bar{D}}$ is a $[\frac{2^m-1}{t},\frac{2^m-1}{t}-m-1,4]$ code.  If there exists a $[\frac{2^m-1}{t},\frac{2^m-1}{t}-m-1]$ binary code with Hamming distance at least $5$, then we have
\begin{equation*}
\begin{split}
\sum_{i=0}^{2}\left(
\begin{array}{cccc}
   \frac{2^m-1}{t}  \\
     i  \\
\end{array}
\right)=1+\frac{2^m-1}{t}+\frac{2^m-1}{2t}\cdot(\frac{2^m-1}{t}-1)>2^{m+1}
\end{split}
\end{equation*}
as $t\neq 2^{\frac{m}{2}}+1$, which is contrary to the sphere packing bound. Hence,
the dual code $(\C(f)^{\bar{D}})^\perp$  is distance-optimal with respect to the sphere packing bound.

\noindent {\bf Case 2:} $3 \, \nmid \, t$. From (\ref{Nab000}) we have
\begin{equation*}
\begin{split}
T(a,b)
&=\frac{1}{3t}\Bigg((-1)^{ {\rm Tr}(a)}\big(\sum_{x \in\mathbb{F}_{2^m}}(-1)^{ {\rm Tr}(bx^{3t})}-1\big)+(-1)^{ {\rm Tr}(a\gamma^{\frac{t(2^m-1)}{3}})}\big(\sum_{x \in\mathbb{F}_{2^m}}(-1)^{ {\rm Tr}(b\gamma^{t}x^{3t})}-1\big)\\
&+(-1)^{ {\rm Tr}(a\gamma^{\frac{2t(2^m-1)}{3}})}\big(\sum_{x \in\mathbb{F}_{2^m}}(-1)^{(b\gamma^{2t}x^{3t})}-1\big)\Bigg).
\end{split}
\end{equation*}
If $b=0$, it is clear that
\begin{equation*}
\begin{split}
T(a,0)
&=\frac{2^m-1}{3t}\left((-1)^{{\rm Tr}(a)}+(-1)^{{\rm Tr}(a\gamma^{\frac{t(2^m-1)}{3}})}+(-1)^{{\rm Tr}(a\gamma^{\frac{2t(2^m-1)}{3}})}\right).
\end{split}
\end{equation*}
From Lemma \ref{eqdad} we have
\begin{equation}\label{Nab0002}
\begin{split}
T(a,0)&=\left\{ \begin{array}{lll}
           \frac{(2^m-1)}{t},  & {\rm occuring} \,\, \, 2^{m-2}\,\, \,{\rm times},\\
           -\frac{(2^m-1)}{3t},
             &  {\rm occuring} \,\, \, 3\cdot2^{m-2}\,\,\, {\rm times}. \end{array}  \right.
\end{split}
\end{equation}
If $b\neq 0$,
then $b$ can be written as $b=\gamma^i$, where $\gamma$ is a primitive element of $\mathbb{F}_{2^m}$ and $1\leq i \leq 2^m-1$. From Lemma \ref{lemeven3} we have
\begin{equation}\label{Nab0003}
\begin{split}
T(a,\gamma^i)=
\begin{cases}
\frac{1}{3t}+\frac{1}{3t}\sum_{x \in\mathbb{F}_{2^m}}\left((-1)^{ {\rm Tr}(\gamma^ix^{3t})}-(-1)^{ {\rm Tr}(\gamma^{i+t}x^{3t})}
-(-1)^{ {\rm Tr}(\gamma^{i+2t}x^{3t})}\right), \\
& \hskip -7cm \text{ if \ ${\rm Tr}(a)=0$, ${\rm Tr}(a\gamma^{\frac{t(2^m-1)}{3}})=1$ and ${\rm Tr}(a\gamma^{\frac{2t(2^m-1)}{3}})=1$,} \\
\frac{1}{3t}+\frac{1}{3t}\sum_{x \in\mathbb{F}_{2^m}}\left(-(-1)^{{\rm Tr}(\gamma^i x^{3t})}+(-1)^{ {\rm Tr}(\gamma^{i+t}x^{3t})}
-(-1)^{{\rm Tr}(\gamma^{i+2t}x^{3t})}\right), \\
& \hskip -7cm \text{ if \ ${\rm Tr}(a)=1$, ${\rm Tr}(a\gamma^{\frac{t(2^m-1)}{3}})=0$ and ${\rm Tr}(a\gamma^{\frac{2t(2^m-1)}{3}})=1$,} \\
\frac{1}{3t}+\frac{1}{3t}\sum_{x \in\mathbb{F}_{2^m}}\left(-(-1)^{ -{\rm Tr}(\gamma^ix^{3t})}-(-1)^{ {\rm Tr}(\gamma^{i+t}x^{3t})}
+(-1)^{{\rm Tr}(\gamma^{i+2t}x^{3t})}\right), \\
& \hskip -7cm \text{ if \ ${\rm Tr}(a)=1$, ${\rm Tr}(a\gamma^{\frac{t(2^m-1)}{3}})=1$ and ${\rm Tr}(a\gamma^{\frac{2t(2^m-1)}{3}})=0$,} \\
-\frac{1}{t}+\frac{1}{3t}\sum_{x \in\mathbb{F}_{2^m}}\left((-1)^{ {\rm Tr}(\gamma^i x^{3t})}+(-1)^{ {\rm Tr}(\gamma^{i+t}x^{3t})}
+(-1)^{{\rm Tr}(\gamma^{i+2t}x^{3t})}\right), \\
& \hskip -7cm \text{ if \ ${\rm Tr}(a)={\rm Tr}(a\gamma^{\frac{t(2^m-1)}{3}})={\rm Tr}(a\gamma^{\frac{2t(2^m-1)}{3}})=0$.} \\
\end{cases}
\end{split}
\end{equation}
Clearly, one of $ 3t \,|\, i$, $ 3t\, |\, (i+t)$ and $ 3t \,|\, (i+2t)$ holds if and only if $t\, |\, i$ for any positive integer $t$ and $1\leq i \leq 2^m-1$. Otherwise, $ 3t \,\nmid\, i$, $ 3t\, \nmid\, (i+t)$ and $ 3t \,\nmid\, (i+2t)$.  Then combining Lemma \ref{lemeven3}, (\ref{Nab0002}) and (\ref{Nab0003}), it is not hard to see that when $(a,b)$ runs over $\mathbb{F}_{2^m}^2$, the value distribution of $T(a,b)$ is given as follows:
\begin{equation}
\begin{split}
T(a,b)&=\left\{ \begin{array}{lll}
\frac{(2^m-1)}{t}, & {\rm occuring} \,\,  2^{m-2}\,\, {\rm times},\\
-\frac{(2^m-1)}{3t},&{\rm occuring} \,\,  3\cdot2^{m-2}\,\, {\rm times},\\
-\frac{1}{t}+\frac{1}{t}\left((t-1)2^{\frac{m}{2}}\right),   &{\rm occuring} \,\,  \frac{(2^m-1)2^{m-2}}{t}\,\, {\rm times},\\
\frac{1}{3t}\left(1-3\cdot2^{\frac{m}{2}}\right), & {\rm occuring} \,\, \frac{(t-1)(2^m-1)2^{m-2}}{t}\,\, {\rm times},\\
\frac{1}{3t}\left(1-(3t-1)2^{\frac{m}{2}}\right),  &{\rm occuring} \,\, \frac{(2^{m}-1)2^{m-1}}{t} \,\, {\rm times},\\
\frac{1}{3t}\left(2^{\frac{m}{2}}+1\right) ,  &{\rm occuring} \,\, \frac{3(t-1)(2^m-1)2^{m-2}}{t}\,\, {\rm times},\\
\frac{1}{3t}\left((3t+1)2^{\frac{m}{2}}+1\right),  &{\rm occuring} \,\, \frac{(2^m-1)2^{m-2}}{t}\,\, {\rm times}.\\
\end{array}  \right.
\end{split}
\end{equation}
By a similar analysis to Case 1, we obtain the weight distribution of $\C(f)^{\bar{D}}$ and the parameters of its dual.
This completes the proof. $\square$

\begin{example}\label{example6}
Let $\C(f)^{\bar{D}}$ be the linear code in Theorem \ref{Theorem5}. Let $m=6$ and $t=3$, then $\C(f)^{\bar{D}}$ has parameters $[21,7,8]$ and its dual has parameters $[21,14,4]$.
The two codes are optimal according to the tables of best codes known in \cite{Grassl2006}.
\end{example}

\begin{remark}
 If $t= 2^{\frac{m}{2}}+1$, it is easy to check that $\C(f)^{\bar{D}}$ is a $[2^{\frac{m}{2}}-1,\frac{m}{2}+1,2^{\frac{m}{2}-1}-1]$ code with the weight enumerator
$$1+(2^{\frac{m}{2}}-1)(x^{2^{\frac{m}{2}-1}-1}+x^{2^{\frac{m}{2}-1}})+x^{2^{\frac{m}{2}}-1},$$ which is optimal with respect to the Griesmer bound. Its dual has parameters $[2^{\frac{m}{2}}-1,\frac{m}{2}+1,4]$,  which is distance-optimal with respect to the sphere packing bound. By the Assmus-Mattson theorem \cite{AM69}, the code $\C(f)^{\bar{D}}$ and its dual support $2$-designs
\cite[Chapter 4]{DingBK18}.  The reader is informed that in the special case $t= 2^{\frac{m}{2}}+1$, the code   $\C(f)^{\bar{D}}$
is permutation-equivalent to a singly punctured code of the first-order Reed-Muller code \cite{LDT20}.
\end{remark}

\vskip 6pt
\subsection{The case that $d(2^k+1)\equiv 2^{\frac{m}{2}}+1 \pmod {2^m-1}$ and $t=2^k+1$}
\vskip 6pt

In this subsection, we always assume that $m$ is even, $d(2^k+1)\equiv 2^{\frac{m}{2}}+1 \pmod {2^m-1}$ and $t=2^k+1$. From (\ref{eq:genc1dode}) it follows that
\begin{equation}\label{Tabq}
\begin{split}
T(a,b)&=\sum_{x \in D}(-1)^{ {\rm Tr}(ax^d+bx)}=\frac{1}{2^{k}+1}\sum_{x \in \mathbb{F}_{2^m}^*}(-1)^{ {\rm Tr}(ax^{2^{m/2}+1}+bx^{2^k+1})} \\
&=\frac{1}{2^{k}+1}\sum_{x \in \mathbb{F}_{2^m}}(-1)^{ {\rm Tr}(ax^{2^{m/2}+1}+bx^{2^k+1})}-\frac{1}{2^{k}+1}.
\end{split}
\end{equation}
 If $k= \frac{m}{2}$, by Lemma \ref{lemeven3}, (\ref{eq:wt12}) and (\ref{Tabq}), $\C(f)^{\bar{D}}$ is a one-weight code with parameters $[2^{\frac{m}{2}}-1,\frac{m}{2},2^{\frac{m}{2}-1}]$, and is permutation-equivalent to a Simplex code.
 In the following, we determine the parameters and the weight distribution of $\C(f)^{\bar{D}}$ and the parameters of
 the dual code $(\C(f)^{\bar{D}})^\perp$ for $k\, \neq\, \frac{m}{2}$.

\begin{theorem}\label{Theorem6}
 Let $d$ satisfy the condition $d(2^k+1)\equiv 2^{\frac{m}{2}}+1 \pmod {2^m-1}$. Let $t=2^k+1$ and $k\, \neq \frac{m}{2}$. Let $\C(f)^{\bar{D}}$ be the linear code defined in (\ref{eq:gencdode}). Then $\C(f)^{\bar{D}}$ is a $[\frac{2^m-1}{t},\frac{3m}{2}, \frac{2^{m-1}-2^{\frac{m}{2}+k-1}}{t}]$ code with the weight distribution in  Table~\ref{Table12}. If $k>1$, its dual has parameters $[\frac{2^m-1}{t},\frac{2^m-1}{t}-\frac{3m}{2},3]$.  If $k=1$ and $m\neq 6$, its dual has parameters $[\frac{2^m-1}{t},\frac{2^m-1}{t}-\frac{3m}{2},4]$,
and is distance-optimal with respect to the sphere packing bound. If $k=1$ and $m=6$, its dual has parameters $[21,12,5]$, and is optimal according to the tables of best codes known in~\cite{Grassl2006}.
\begin{table}[h]
\caption{\rm Weight distribution of $\C(f)^{\bar{D}}$ in Theorem \ref{Theorem6} }\label{Table12}
\begin{center}
\begin{tabular}{cccc}\hline
Weight      & Multiplicity \\\hline
$0$ &  $1$ \\   \hline
  $\frac{2^{m-1}+2^{\frac{m}{2}-1}}{t}$ & $\frac{2^{3k}(2^{\frac{m}{2}}-1)(2^m-2^{m-2k}-2^{m-3k}+2^{\frac{m}{2}}-2^{\frac{m}{2}-k}+1)}{t^2(2^{k}-1)}$  \\   \hline
 $\frac{2^{m-1}-2^{\frac{m}{2}+k-1}}{t}$ & $\frac{2^k(2^m-1)(2^{\frac{m}{2}}+2^{\frac{m}{2}-k}+2^{\frac{m}{2}-2k}+1)}{t^2}$  \\   \hline
$\frac{2^{m-1}+2^{\frac{m}{2}+2k-1}}{t}$ & $\frac{(2^{\frac{m}{2}-k}-1)(2^m-1)}{t^2(2^{k}-1)}$  \\   \hline
\end{tabular}
\end{center}
\end{table}
\end{theorem}
{\it Proof.} It is clear that ${\rm Tr}\left(ax^{2^{m/2}+1}\right)={\rm Tr}_1^{\frac{m}{2}}\left((a+a^{2^{m/2}})x^{2^{m/2}+1}\right)$ and $a+a^{2^{m/2}} \in \mathbb{F}_{2^{\frac{m}{2}}}$ for any $a \in \mathbb{F}_{2^m}$. Obviously, $a+a^{2^{m/2}}$ runs through $\mathbb{F}_{2^{\frac{m}{2}}}$ with multiplicity $2^{\frac{m}{2}}$ when $a$ runs through $\mathbb{F}_{2^m}$. Let
$$K=\left\{ x\in \mathbb{F}_{2^m} : x+x^{2^{m/2}}\right\}.$$  Then $\mathbf{c}(a,b)=\mathbf{c}(a,b+\delta)$ for any $\delta \in K$.
Since $t \, | \, (2^m-1)$ and $t=2^k+1$, it is easy to prove that there exists a positive integer $\ell$ such that $\ell (2^k+1)\equiv 2^{\frac{m}{2}}+1 \pmod {2^m-1}$ if and only if $v_2(k)=v_2(\frac{m}{2})$ and $k\, | \, \frac{m}{2}$. From Lemma \ref{lemeveeen3}, (\ref{eq:wt12}) and (\ref{Tabq}), the desired weight distribution of $\C(f)^{\bar{D}}$ follows.

Let  $A^{\perp}_{i}$ be the number of the codewords with weight $i$ in $(\C(f)^{\bar{D}})^\perp$. By the first four  Pless
power moments, we get that $A_1^{\perp}=A_2^{\perp}=0$ and
\begin{equation*}
\begin{split}
A_3^{\perp}&=\frac{1}{48(2^k+1)^5(2^{2k}-1)}\big((2^{7k}+2^{6k+1}+6\cdot 2^{2k}+4\cdot 2^{3k}+7 \cdot 2^k+2-3 \cdot 2^{5k}-10 \cdot 2^{4k}-9 \cdot 2^{3k})\cdot 2^{\frac{3m}{2}}\\
&+(3\cdot 2^{5k}+10 \cdot 2^{4k}+5\cdot 2^{3k}-6\cdot 2^{2k}-7\cdot 2^k-2)\cdot 2^{\frac{5m}{2}}\big).
\end{split}
\end{equation*}
We can check that $A_3=0$ if $k=1$ and $A_3\neq0$ if $k>1$. If $k=1$, by the fifth  Pless power moment, we obtain that
 \begin{equation*}
\begin{split}
A_4^{\perp}=\frac{1}{6^4}(2^{4m}+70\cdot 2^{\frac{5m}{2}}-6\cdot 2^{\frac{7m}{2}}-25\cdot 2^{3m})+\frac{2^{2m}}{54}-\frac{2^{\frac{3m}{2}+2}}{81}.
\end{split}
\end{equation*}
It is easy to check that $A_4^{\perp}=0$ if and only if $m=6$. Similarly to the proof of Theorem \ref{Theorem5}, we can
show that $(\C(f)^{\bar{D}})^\perp$  is distance-optimal with respect to the sphere packing bound if $k=1$ and $m\neq 6$.
By the sixth  Pless power moment, we obtain $A_5^{\perp}\neq 0$. This completes the proof. $\square$

\begin{example}\label{example7}
Let $\C(f)^{\bar{D}}$ be the linear code in Theorem \ref{Theorem6}. Let $m=10$ and $k=1$. Then $\C(f)^{\bar{D}}$ has parameters $[341,15,160]$. Its dual has parameters $[341,326,4]$ and
is distance-optimal with respect to the sphere packing bound.
\end{example}

We settled the parameters and weight distribution of the code $\C(f)^{\bar{D}}$ in Theorem \ref{Theorem6}, but do not
know if the corresponding code $\C(f)$ was studied in the literature or not.

\section{Summary and concluding remarks }\label{sec:concluding}

The main contributions of this paper are the following:
\begin{enumerate}
\item We obtained several classes of binary punctured codes with three weights, or four weights,
or five weights, or six weights, and determined their weight distributions (see Theorem \ref{Theorem1}, Corollary \ref{coroll}, Theorem \ref{Theorem2}, Theorem \ref{Theorem3}, Theorem \ref{Theorem4}, Theorem \ref{Theorem5} and Theorem \ref{Theorem6}).

\item  We presented several classes of self-complementary linear codes. Almost all of their duals are
distance-optimal with respect to the sphere packing bound (see Theorem \ref{Theorem1}, Corollary \ref{coroll}, Theorem \ref{Theorem2}, Theorem \ref{Theorem3}, Theorem \ref{Theorem4}, Theorem \ref{Theorem5} and Theorem \ref{Theorem6}).

\item  We got some distance-optimal codes with specific parameters (see Example \ref{example1}, Example \ref{example2}, Example \ref{example5}, Example \ref{example3} and Example \ref{example6}).

\end{enumerate}
A constructed binary linear code $\C$ is new if one of the following happens:
\begin{itemize}
\item No binary linear code with the same parameters was documented in the literature.
\item Some binary linear codes with the same parameters as $\C$ were documented in the literature, but their weight distributions are different from
          the weight distribution of $\C$.
\item Some binary linear codes with the same parameters and weight distribution as those of $\C$ were documented
          in the literature, but they are not
          permutation-equivalent to $\C$.
\end{itemize}
Except the class of codes in Remark 5.4, every other class of binary codes presented in this paper would be new, as we have not found  a class of binary codes with the same parameters and weight distributions  in the literature as those codes documented in this paper.

Starting from Section \ref{sec-hubei2}, we restricted our discussions on finite fields with characteristic 2. The position sets were constructed from some trace functions and cyclotomic classes. It would be interesting to extend some of the results in this paper to the case that $q \geq 3$.

Finally, we make some comments on the puncturing and shortening techniques. As mentioned in the introductory section,
every projective linear code over $\bF_q$ is a punctured Simplex code over $\bF_q$ and a shortened code of a Hamming code over $\bF_q$. However, it is in general very hard to determine the parameters of punctured codes of Simplex codes and shortened codes of Hamming codes \cite{LDT20, XTD20}. Hence, we still need to study punctured and shortened codes of
other families of linear codes.  For a given linear code $\C$ and a set $T$ of coordinate positions in $\C$, it may be possible to
determine the parameters of the punctured code $\C^T$ when $|T|$ is small, but it is very hard to do so in general if $|T|$ is large \cite{TDX20}.

\end{document}